\pdfoutput=1

\documentclass[11pt]{article}
\usepackage{jheppub}

\usepackage{caption}
\usepackage{extarrows}

\usepackage{mathrsfs}

\usepackage{amsmath}
\usepackage{latexsym,amssymb}
\usepackage{graphicx,color,slashed}
\usepackage{bbm}
\usepackage{mathtools}
\usepackage{ifpdf}
\newcommand{\ft}[2]{{\frac{#1}{#2}}}

\usepackage{amsfonts}
\usepackage{amsmath}
\usepackage{latexsym,amssymb}
  \usepackage{amsmath,amssymb}%,amsthm}
\DeclareSymbolFontAlphabet{\Scr}{rsfs}

\newcommand{\cG}{\mathcal{G}}

\newcommand{\cA}{\mathcal{A}}
\newcommand{\cB}{\mathcal{B}}

\newcommand{\cH}{\mathcal{H}}
\newcommand{\cL}{\mathcal{L}}
\newcommand{\cM}{\mathcal{M}}
\newcommand{\cN}{\mathcal{N}}

\newcommand{\cV}{\mathcal{V}}
\newcommand{\Tr}{\mathrm{Tr\,}}

\newcommand{\be}{\begin{equation}}
\newcommand{\ee}{\end{equation}}
\newcommand{\ba}{\begin{eqnarray}}
\newcommand{\ea}{\end{eqnarray}}

\renewcommand{\d}{\textrm{d}}

\renewcommand{\a}{\alpha}

\newcommand{\N}{\mathcal{N}}

\def\rmi{{\rm i}}

\def\E{{$E_{7(7)}$}}
\def\EE {{$E_{5(5)}$}}
\def\adot{{\dot a}}
\def\c{\gamma}
\def\I{\mathcal{I}}

\renewcommand{\l}{\lambda}

\newcommand{\rf}[1]{(\ref{#1})}
\newcommand{\bea}{\begin{eqnarray}}
\newcommand{\eea}{\end{eqnarray}}

\def\bfzero{\relax{\rm I\kern-.18em 0}}
\def\bfone{\relax{\rm 1\kern-.35em 1}}
\def\twomat#1#2#3#4{\left(\begin{array}{cc}
\end{array}
\right)}

\def\d{\delta}

\def\cH{{\cal H}}

\newcommand{\C}{\mathds{C}}

\newcommand{\SO}{\mathop{\rm SO}}
\newsavebox{\uuunit}
\sbox{\uuunit}
    {\setlength{\unitlength}{0.825em}
     \begin{picture}(0.6,0.7)
        \thinlines
        \put(0,0){\line(1,0){0.5}}
        \put(0.15,0){\line(0,1){0.7}}
        \put(0.35,0){\line(0,1){0.8}}
       \multiput(0.3,0.8)(-0.04,-0.02){12}{\rule{0.5pt}{0.5pt}}
     \end {picture}}
\newcommand {\unity}{\mathord{\!\usebox{\uuunit}}}
\newcommand{\Gammabig}{\tilde{\mathbf{\Gamma}}}

\newcommand{\fe}{\mathfrak{e}}
\newcommand{\fp}{\mathfrak{p}}
\newcommand{\fso}{\mathfrak{so}}

\definecolor{hs}{rgb}{0.87, 0.0, 0.0}

\title{\rm{\bf      Gauge-fixing local H symmetry in supergravities }}
\author[a]{ Renata Kallosh,}
\author[b] {Henning Samtleben,}
\author[c,d] {and Antoine Van Proeyen}
 \affiliation[a] {Stanford Institute for Theoretical Physics and Department of Physics,\\
Stanford University, Stanford, CA 94305, USA}
\affiliation[b] {ENSL, CNRS, Laboratoire de Physique, F-69342 Lyon, France \\
Institut Universitaire de France (IUF)}
 \affiliation[c] {Institute for Theoretical Physics, KU Leuven, Celestijnenlaan 200D, B-3001 Leuven, Belgium}
 \affiliation[d] {Leuven Gravity Institute, KU Leuven, Celestijnenlaan 200D, box 2415, B-3001 Leuven, Belgium}
\emailAdd{kallosh@stanford.edu}
\emailAdd{henning.samtleben@ens-lyon.fr}
\emailAdd{antoine.vanproeyen@kuleuven.be}
\abstract{  We discuss known maximal D-dimensional supergravities of  two types:  type I with G/H coset spaces and  type II derived by compactification from higher dimensions without dualization, these have less manifest symmetries. In  4D and 6D in type I models
 we perform explicit gauge-fixing of local H symmetries  in  unitary gauges: symmetric,  Iwasawa and partial Iwasawa. In 4D supergravity I in symmetric gauge  global H-invariance and nonlinearly realized G-symmetry are valid on shell, classically.   The global H-symmetry and  G-symmetry in Iwasawa-type gauges in type I and  in type II supergravities are not manifest,  if at all present. This fact raises the issue of the gauge equivalence of the  S-matrix of various gauge-fixed D-dimensional supergravities and  its relation to the ones computable using superamplitude methods.}

\begin{document}

\maketitle

 %\tableofcontents{}

\newpage

\parskip 5pt

%%%%%%%%%%%%%%%%%%%%%%%%%%%%%%%%%%%%%%%%%%%%%%%%%%%%%%%%%%%%%%%%%%%%%%
%%%%%%%%%%%%%%%%%%%%%%%%%%%%%%%%%%%%%%%%%%%%%%%%%%%%%%%%%%%%%%%%%%%%%%
%%%%%%%%%%%%%%%%%%%%%%%%%%%%%%%%%%%%%%%%%%%%%%%%%%%%%%%%%%%%%%%%%%%%%%
%%%%%%%%%%%%%%%%%%%%%%%%%%%%%%%%%%%%%%%%%%%%%%%%%%%%%%%%%%%%%%%%%%%%%%
%%%%%%%%%%%%%%%%%%%%%%%%%%%%%%%%%%%%%%%%%%%%%%%%%%%%%%%%%%%%%%%%%%%%%%
\section{Introduction}
Consider well known  (ungauged) maximal  supergravity theories in 4D and 6D. They   contain physical scalar fields which belong to the coset space  G/H.  Here G  is a non compact group, while H is the maximal compact subgroup of G:
\be
{G\over H}\Big |_{4}= {E_{7(7)}\over SU(8)} \, , \qquad \qquad {G\over H}\Big |_{6}= {E_{5(5)}\over USp(4)\times USp(4) }
\ee
In the  original versions of ungauged supergravities in \cite{Cremmer:1979up,deWit:1982bul,Tanii:1984zk,Bergshoeff:2007ef} the number of scalars is defined by
the fundamental representation of a group G, and  there is also a local H symmetry, which is the $R$-symmetry group. We will refer to these supergravities as supergravities of type I. In all integer dimensions D these supergravities of type I have physical scalars in the  coset space  $(G/H)_D$ and they have the corresponding global $G_D$ and local $H_D$ symmetries.

D-dimensional supergravities of the type I have global  U-duality symmetry $G_D$. These are  groups $E_{11-D(11-D)}$ \cite{Hawking:1981bu,Hull:1994ys}, they are often called $E_{11-D}$ groups.
 They are nicely explained by B. Julia \cite{Hawking:1981bu} in terms of Dynkin diagrams as the process of  Group Disintegration.   Starting from $E_{8(8)}$ in 3D removing the right node of the Dynkin diagram, one by one, one  gets $E_{7(7)}$ in 4D, $E_{6(6)}$ in 5D, $E_{5(5)}$ in 6D,  and all the way up.

The local H symmetry in \cite{Cremmer:1979up,deWit:1982bul,Tanii:1984zk,Bergshoeff:2007ef} has unusual features.
In  standard local gauge symmetries there are propagating gauge fields, but in local H symmetries the role of the gauge field is played by the composite scalar dependent connection. Some of the  advantages in keeping  local H symmetry not gauge fixed is that the global duality  symmetry G and the local H symmetry are independent, and fermions transform under H and are neutral in G.

The action with local H symmetry depends on scalars which are in the adjoint  representation of G,  for example 133 in maximal 4D supergravity
and  45  in 6D.
These scalars  parametrize a G-valued matrix $\cV(x)$ which transforms by G from the left and by H from the right.
When  local H symmetry in supergravities is gauge-fixed, only physical  scalars remain, their number is reduced to the number of coordinates in the coset space G/H, 70 in 4D and 25 in 6D. In symmetric gauges  fermions in gauge-fixed theory  transform under G symmetry due to a compensating H symmetry transformation, preserving the choice of the gauge. In 4D case symmetric gauge  in \cite{Cremmer:1979up,deWit:1982bul} supergravity   was  studied in detail in \cite{Cremmer:1979up,deWit:1982bul,Kallosh:2008ic,Bossard:2010dq}, where it was shown that the global H-symmetry is valid on shell.

Thus,  it is convenient to refer to  original versions of ungauged supergravities in D-dimensions,  with global G and local H symmetries, as supergravities I. We will refer to supergravities derived by compactification from (D+n) dimensional supergravities, {\it without dualization}, as supergravities II. These are less known, in general, however, these models have played an important role in studies of black hole attractors in maximal 4D supergravity, see for example \cite{Ferrara:2006em,Ferrara:1997ci,Ferrara:1997uz,Andrianopoli:2002mf,Ceresole:2009jc}. Specifically, 1/2, 1/4, 1/8 extremal BPS black holes are associated with type I supergravity,  but the non-BPS Kaluza-Klein extremal black holes are associated with type II supergravity \cite{Andrianopoli:2002mf,Ceresole:2009jc}.

We will mostly focus on the case of $n=1$ when D-dimensional supergravity is derived from the D+1 supergravity compactified on a circle. In 4D these type II supergravity are models in  \cite{Andrianopoli:2002mf,Ceresole:2009jc} and in 6D in \cite{Cowdall:1998rs}.

Our study of gauge-fixing H-symmetry in  supergravities is based on the original work  in 4D and in 6D in \cite{Cremmer:1979up,deWit:1982bul,Tanii:1984zk,Bergshoeff:2007ef} as well as on
  supergravity studies in diverse dimensions in  \cite{Salam:1989ihk,Andrianopoli:1996ve,Andrianopoli:1996zg,Cremmer:1997ct,Tanii:1998px, deWit:2002vz,Samtleben:2008pe,Trigiante:2016mnt,Sezgin:2023hkc}.  We will define classes of gauges we are interested  in various dimensions, and we will fill in some gaps in the existing literature, for example in gauge-fixing of 6D maximal supergravity,  as well as with regard to Iwasawa-type gauges in various dimensions.

{\it   Unitary gauges in the context of quantum field theory} have the property that the gauge fixing function depends on the fields and not on their derivatives. Therefore there are no Faddeev-Popov propagating  ghost fields.

Unitary {\it symmetric gauges} are well known in 4D, but not in 6D. In maximal 4D supergravity after gauge-fixing  local $H= SU(8)$ symmetry \cite{Cremmer:1979up,deWit:1982bul}, there is a remaining nonlinearly realized G=\E\, and   a  field-dependent compensating H= SU(8)  symmetry. The physical scalars $\phi_{ijkl}, \, \bar \phi^{pqmn}$ transform in a linear representation of $SU(8)$. All dependence on physical scalars is non-polynomial. The 1-loop anomalies of global $SU(8)$-symmetry in this gauge cancel \cite{Marcus:1985yy}.

Unitary  {\it Iwasawa gauges}, also called triangular gauges,  in supergravities were described already in \cite{Cremmer:1979up}.  These gauges  have a remarkable polynomiality in some of the scalars, which is absent in symmetric gauges. These are associated in case of maximal supergravities with dimensional reduction from 11D supergravity. In these gauges there is a  maximal number of  axionic scalars, which enter the action polynomially.

Unitary {\it partial Iwasawa gauges} we define here as   the ones  where D-dimensional theories  are  associated with dimensionally reduced maximal   $(D+1)$ supergravities. They have the feature that the  D+1 coset $(G/H)_{D+1}$ inherited in dimension D  was gauge-fixed in a symmetric gauge in D+1. The number of axionic polynomial scalars in these gauges is always non-zero, but less than the one in Iwasawa triangular gauge.

Supergravity actions depend on  scalars via the vielbein $\cV(x)$.
The vielbein transforms under global $G$ symmetry and local H-symmetry
\be
\cV(x) \to {\bf g}\, \cV(x)h^{-1}(x)
\ee
Before gauge-fixing the vielbein is in a  fundamental representation of G, the number of scalars is dim [G]. After gauge-fixing
\be
\cV(x)\to \cV(x)_{g.f.}\, .
\ee
It is a matrix depending only on  physical scalars, where the number of physical scalars $n_{sc}$ is equal to
dim [G] - dim [H].

 Different choices of coset representatives define different choices of gauges of local H symmetry. Various solutions of underlying mathematical problem to find    G/H coset space representatives can be found in supergravity original papers \cite{Cremmer:1979up,deWit:1982bul,Tanii:1984zk,Bergshoeff:2007ef}  and  reviews in \cite{Andrianopoli:1996ve,Andrianopoli:1996zg,Cremmer:1997ct,Tanii:1998px, deWit:2002vz,Samtleben:2008pe,Trigiante:2016mnt,Sezgin:2023hkc} and
 in the textbooks like \cite{Gilmore:2008zz}, \cite{Helgasson}.

\begin{figure}
\centering
\includegraphics[scale=0.6]{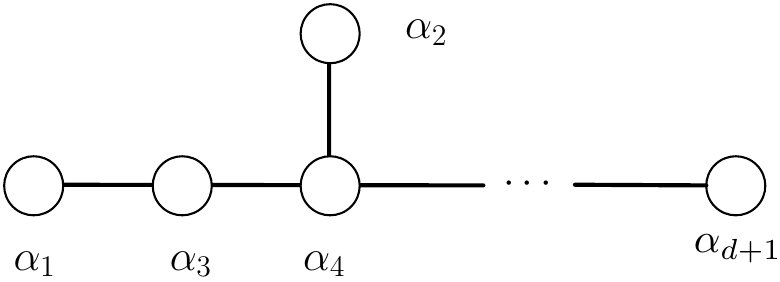}
\caption{\footnotesize  The  Dynkin diagrams of U-duality groups $E_{d+1(d+1)}$ as given in \cite{Green:2010kv}, $0\leq d\leq 7$, $D=10-d$. The groups $E_{d+1(d+1)}$ are often called $E_{d+1}$.
}
\label{dynkin}
\end{figure}

Consider   Fig.\,\ref{dynkin} here for Dynkin diagrams of U-duality groups as presented in \cite{Green:2010kv}.
Different gauges for local H-symmetry correspond to different ways of cutting some nodes in these Dynkin diagrams, thereby breaking  global G-symmetry of the classical action in a class of Iwasawa type gauges.

The purpose of this paper is to provide a systematic  gauge-fixing of local H-symmetries in D-dimensional supergravities in a class of gauges which either preserve global H-symmetry and G-symmetry, in symmetric gauges, or not (at least not manifestly), in  Iwasawa-type gauges associated with D+1 supergravities. These various choices of gauge-fixing can be studied in the future to understand how the existence of different unitary gauges might affect the quantization of supergravities, the issue of local H-symmetry and global G-symmetry anomalies and UV divergences.

In particular, the 1-loop anomalies of global H-symmetry  were computed in  \cite{Marcus:1985yy} in a symmetric gauge, where global H-symmetry is present in a classical action and where the vielbein is assumed to take a form
\be
\cV_{g.f.} = e^{\phi\cdot \mathbb{K}}
\ee
Here $\mathbb{K}$ are generators  in a noncompact part of the algebra of G, which is a property of  symmetric gauges. In other gauges  the status of a global H-symmetry is not obvious, unless one can prove the on shell gauge independence of these theories. Therefore the issue of 1-loop anomalies in supergravity  with account of  symmetric and  Iwasawa-type  gauges requires an additional investigation.

\section{D-dimensional  supergravity I and supergravity II   }
\begin{figure}
\centering
\includegraphics[scale=0.35]{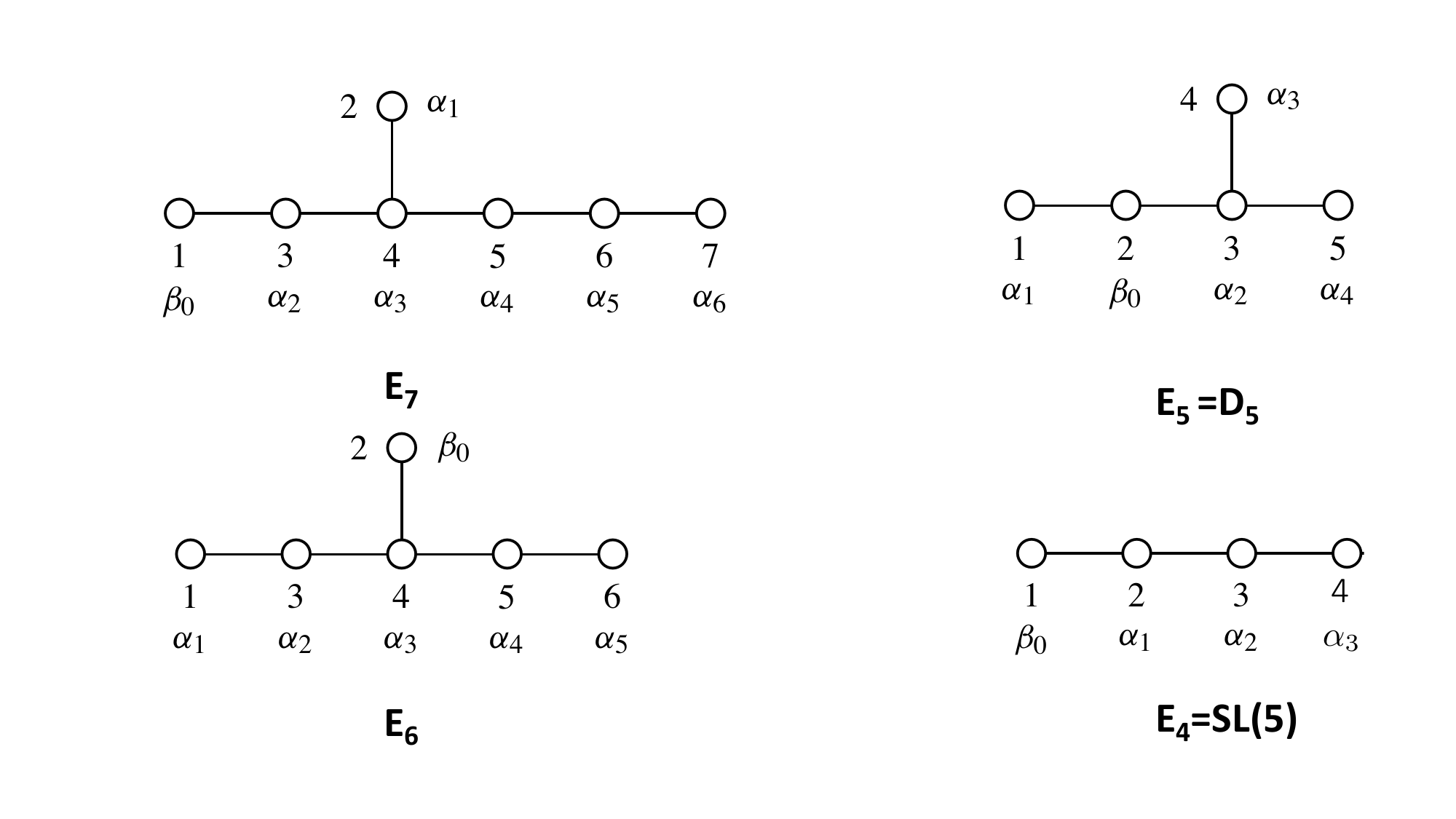}
\caption{\footnotesize  Left panel at the top shows  Dynkin diagram for $E_7$, the one on the bottom is $E_6$. Right panel at the top shows Dynkin diagram for $E_5$, the one on the bottom is $E_4=SL(5)$.
Figures taken from \cite{Kazhdan:2001nx} where a detailed explanation can be found. Deleting the right node in $E_7$ we see  $E_6$ and deleting the right node in $D_5=$ \EE\, we see $SL(5)= E_{4(4)}$. }
\label{EA}
\end{figure}

Supergravities I were constructed in dimension D, the physical scalars are in $(G/H)_D$ coset space. These models have local H-symmetry and global on shell G symmetry.
This local H symmetry can be gauge-fixed in various gauges. In symmetric gauges there is a  non-polynomial dependence on all physical scalars and on shell $E_{11-D(11-D)}$ symmetry.  Upon gauge-fixing the global G symmetry is still preserved but non-linearly realized. In Iwasawa and partial Iwasawa gauges there is a polynomial dependence on some of the  physical scalars, often called axions, and  non-polynomial dependence on other  physical scalars, often called dilatons.

In Iwasawa type  gauges the status of a global $H_D$ symmetry and on shell $G_D$-symmetry is  not known, a priory. The symmetries, like \E\, in 4D and \EE\, symmetry in 6D are  broken off shell, by construction. This is best explained using the
Dynkin diagrams in Fig. \ref{dynkin} and Fig. \ref{EA}.

We will use two of the parabolic subgroups of the U-duality groups described in \cite{Green:2010kv}. The first one is  the subgroup $P_{\alpha_{d+1}}$ obtained by removing the root $\alpha_{d+1}$  associated with
the most right last node $d + 1$ of the Dynkin diagram in Fig. \ref{dynkin}. These gauges in D-dimensional supergravity are related to actions derived  by compactification from D+1 supergravity in accordance with the Group Disintegration \cite{Hawking:1981bu}. The second subgroup $P_{\alpha_{2}}$ is obtained by removing the root $\alpha_{2}$  associated to the
node 2. These are triangular, Iwasawa  unitary gauges for local H-symmetry, they  were explored in   \cite{Cremmer:1997ct}  for D-dimensional supergravities derived  by compactification  of 11D supergravity on a torus
$ T^{d+1}$, d=10-D.

In Fig. \ref{EA} we show examples of Iwasawa gauges where right node was removed from the Dynkin diagram: in 4D case this describes the choice of the gauge in type I supergravity where \E\, group is broken down to $E_{6(6)}$. In 6D the \EE\, duality is broken to $E_{4(4)}$.

Supergravities II in dimension D which we focus on  are related to (D+1)-dimensional supergravity. They were derived from D+1 supergravities by a reduction on a circle, or, sometimes  were constructed directly in dimension D.
These correspond to cases of parabolic subgroups $P_{\alpha_{d+1}}$ of $E_{d+1(d+1)}$ group with removal of the right node of a Dynkin diagram in Fig. \ref{dynkin}.
Their local symmetry is smaller, it is at most $H_{D+1}$-symmetry inherited from D+1. This local $H_{D+1}$-symmetry can be gauge fixed already in D+1, before dimensional reduction, or after, see for example
  \cite{Andrianopoli:2002mf,Cowdall:1998rs}.

Supergravities II are also given by their actions in dimension D and they also have the same amount of maximal local supersymmetries as supergravities I. These models in general case  were described in \cite{Andrianopoli:1996zg}.
The physical scalars correspond to a decomposition of the $(G/H)_D$ coset space via the smaller coset space of D+1 supergravity $\frac{G_{D+1}}{H_{D+1}}$, the radius of the circle, $r_{D+1}$ and the scalars originating from the extra component of D+1 vectors ${\bf V}_r^{D+1}$.
\be
\frac{G_D}{H_D} \sim \Big ( \frac{G_{D+1}}{H_{D+1}}, r_{D+1}, {\bf V}_r^{D+1}\Big)
\label{cosetideal}
\ee
It means that supergravities II always have some axions, as opposite to supergravities I where there is a class of symmetric gauges without axions. In supergravities II the minimal value of the axions is reached in a gauge where the local $H_{D+1}$ symmetry is gauge-fixed in a symmetric gauge.

 In supergravities I the number of axions depends on the gauge chosen to fix the local H-symmetry. It varies between zero and maximum in Iwasawa triangular gauge.

\noindent 1. In symmetric gauge, there are no polynomial axions, all scalars are non-polynomial.
 \be
\# {\rm axions}_{sym} =0
 \ee
2. The Iwasawa triangular gauges  are closely related to the supergravity actions derived by dimensional reduction from 11D supergravity. They
involve a Borel subalgebra of the algebra of  the group G, or they can also be described in terms of the solvable Lie group and using
 the solvable parametrization of the coset space G/H \cite{Samtleben:2008pe,Trigiante:2016mnt}.   The number of axions in these gauges in dimension D is defined by the properties of the U-duality group $G= E_{11-D (11-D)}$  in a given dimension D
 \be
\# \, {\rm axions}_{triangular} ={1\over 2} ({\rm dim} \, G- {\rm rank} \, G)
 \ee
3.  In gauges we call partial Iwasawa gauges related to D+1 supergravities the scalars are defined by the decomposition of the coset space $(G/H)_D$  in eq. \rf{cosetideal}. In these gauges of supergravity I, as in supergravities II,
there are  axions related to
 compactified vectors ${\bf V}_r^{D+1} $ presenting  the abelian ideal in $D$ dimensions.  The
 number of axions is the number of vector fields in one dimension above, it also has a group theoretic interpretation as a maximal abelian ideal \cite{Andrianopoli:1996zg}
\be
\# {\rm axions}_{partial \, \, Iwasawa} ={\rm dim} \, {\cA}_D
 \ee
 This is the case when  the relevant H-symmetry in D+1 theory was gauge-fixed in  symmetric  gauge. However, when the gauge in D+1 supergravity was an Iwasawa gauge, the number of axions is exceeding the ones in partial Iwasawa gauge.

 To summarize, we find that in supergravity I there are gauges which have the same set of scalars as supergravity II, derived from D+1 supergravity by compactification on a circle. If in supergravity II a local $H_{D+1}$ symmetry was gauge-fixed in  Iwasawa gauge, the corresponding supergravity I needs to be gauge-fixed in the appropriate Iwasawa gauge. If  in supergravity II a local $H_{D+1}$ symmetry was gauge-fixed  in  a symmetric gauge, supergravity I has to be gauge-fixed in an appropriate  partial Iwasawa gauge, to reproduce the same set of scalars. In general, the relation between vectors in supergravities I and II in various gauges is not as simple.

Thus, using different gauges in supergravities I we can study both supergravities I as well as supergravities II. The minimal number of axions in supergravities II is the same as in supergravities I in the partial Iwasawa gauge and always positive.
\be
\# {\rm axions}_{partial \, \, Iwasawa} = \# {\rm axions}_{supergravity\,  II} = {\rm dim} \, {\cA}_D
>0
\ee
This is opposite to symmetric gauges of supergravity I
 \be
\# {\rm axions}_{sym} =\# {\rm axions}_{supergravity \, I} =0
 \ee
  The mere existence of supergravities II in \cite{Andrianopoli:2002mf} and  in  \cite{Cowdall:1998rs} as well as
 the fact that there are gauges in supergravities I where scalars are the same as  in supergravities  II, suggest that these issues might be relevant to quantum properties of maximal supergravities.

In supergravity I the G-valued matrix, a vielbein  $\cV(x)$, transforms by G from the left  and by H from the right. Before local H-symmetry is gauge-fixed, these are independent transformations
\be
\cV(x) \to {\bf g}\, \cV(x)\, , \qquad  \cV(x) \to  \cV(x) h^{-1}(x)\, , \qquad {\bf g}\in G\, ,  \qquad h(x) \in H
\label{old1}\ee
The Lie algebra $\mathfrak{g}$ of a group G can be decomposed into two orthogonal subspaces: the Lie algebra
 $\mathfrak{h}$ of a group H and a coset space $\mathfrak{k}$. Here  H is the maximal compact group in G.
\be \label{GK}
\mathfrak{g}= \mathfrak{h} \oplus\mathfrak{k}
 \qquad [\mathfrak{h}, \mathfrak{h}] \subset \mathfrak{h}\,  ; \qquad [ \mathfrak{h},\mathfrak{k}] \subset  \mathfrak{k}\,   ; \qquad[ \mathfrak{k}, \mathfrak{k}] \subset \mathfrak{h}
\ee
The  supergravity reviews   which we will use here have the following features. In  \cite{Tanii:1998px, deWit:2002vz, Samtleben:2008pe} one starts with ungauged supergravity with a local H symmetry and global G symmetry, both independent and linearly realized. The action depends on scalars which form a fundamental representation of G. These scalars  parametrize a G-valued matrix $\cV(x)$ which transforms by G and by H as shown in eq. \rf{old1}.
Gauge-fixing local H symmetry in unitary gauges requires to use all parameters of  local  H symmetry, like 63 in $H=SU(8)$ in 4D, or 20 in $H=SO(5)\times SO(5)$ in 6D,  to bring the theory to the form where it depends only on physical scalars.  Their number is dim [G]-dim [H], i. e. in maximal 4D we have 133-63=70  and in 6D 45-20=25  physical scalars.

Our  purpose is, in particular, to compare supergravity actions derived in a given dimension D=4,\,6 versus the ones obtained via dimensional reduction from D+1=5,\,7. For this purpose it is convenient to study all theories in dimension D at the stage where local H symmetry was not yet gauge-fixed. It means that we study supergravities I and gauge-fix them in various gauges.

In such case\footnote{In \cite{Andrianopoli:1996ve,Trigiante:2016mnt}  the stage of an ungauged supergravity with a local H symmetry is replaced from the beginning by various  choices of coset space representatives $L(\phi^r)$. Namely, the coset representatives depend only on physical scalars $\phi^r, r=1,\dots , n_{sc}$. There is no local H symmetry independent on G symmetry, instead the action of G transforms the fields $\phi^r$ in a coset between themselves and might require a compensating field-dependent H transformation.
} we define these gauges in terms of the G-valued vielbein matrix $\cV(x)$.
The corresponding gauge-fixings which eliminate unphysical scalars correspond to certain choices of the vielbein $\cV$: for example, one gauge is a symmetric one, the other is an  Iwasawa  gauge, or  a partial Iwasawa gauge.  All of these are unitary gauges, in a sense that they have only physical scalars and do not require propagating  ghost fields.

\section{Symmetric,  Iwasawa  and partial Iwasawa supergravity I unitary gauges   }

 In D-dimensional supergravities with a local H symmetry, i. e. in supergravities I, one can make a choice of
the vielbeins/of the coset representatives of a different kind. We discuss   possible choices of unitary gauges starting with the well known {\it symmetric gauges} \cite{Cremmer:1979up,deWit:1982bul,Sezgin:1981ac}.

There are also gauges suitable for addressing the relation between D-dimensional supergravity with $(G/H)_D$ coset space with the one in D  derived from higher dimensions, from 11d or from D+1, for example.
They gauges are, in general,  different. But all of them have some axionic scalars interacting polynomially.

{\it Iwasawa triangular gauges} are also well known, they represent 11D supergravity dimensionally reduced to D-dimension, see \cite{Cremmer:1979up,Cremmer:1997ct,Samtleben:2008pe}.
There is also  a class of gauges  available in the literature, for example in \cite{Sezgin:1981ac}, where D-dimensional supergravity was derived from a D+1-dimensional supergravity by compactification on a circle. If the relevant $H_{D+1}$-symmetry in D+1 theory was gauge-fixed in a symmetric gauge,  this theory can be related to D-dimensional supergravity in a class of gauges which we call   {\it  partially  Iwasawa  gauges}.

The interest to parabolic subgroups of the U-duality groups in string theory and supergravity was raised in the context of automorphism analysis  in \cite{Obers:1999um,Green:2010wi}. It was shown that  maximal parabolic subgroups of $E_{d(d)}$ (often called $E_d$)  arising in string theory are of the form $GL(1)\times X_{d-1}$. For example,  for $E_{7(7)}$ the $X_{6}$  is  $E_{6(6)}$  and for  \EE\,  the $X_{5}$ is $SL(5)$.
This is convenient to see in cases interesting for us using Dynkin diagrams in Fig. \ref{EA} here, which we took from \cite{Kazhdan:2001nx}. In particular, after deleting the right node $E_7$ becomes $E_6$ and $D_5=$ \EE\, becomes $SL(5)=A_{4}\sim E_4$, see also table 2 in  \cite{Green:2010wi}.

The reason for  Iwasawa   and partial Iwasawa gauges  is the fact that in dimension D+1, the scalars are in a smaller coset $(G/H)_{D+1}$, see eq. \rf{cosetideal}.
These are gauges suitable for addressing the relation between D-dimensional supergravity with $(G/H)_D$ coset space with the one in D  derived from higher dimensions, they are, in general,  different. The ones in dimension D derived from D+1 dimensions  have some axionic type scalars where, according to eq. \rf{cosetideal}, the axions correspond to ${\bf V}_r^{D+1}$. The local symmetry in D+1 is also smaller, it is $H_{D+1}$. One can gauge-fix  $H_{D+1}$ either  in a the triangular, Iwasawa gauge or in  a symmetric gauge. In dimension D  the 1st choice will produce a triangular Iwasawa  gauge,
the second choice will produce partial Iwasawa gauge. The simplest example of this partial Iwasawa  gauge one can see in the 4D maximal supergravity action in  \cite{Andrianopoli:2002mf}. This is a limit of the action in  \cite{Sezgin:1981ac} when all mass/gauging parameters $m_i \to 0$.

\begin{itemize}

\item {\it Symmetric gauge}

 The reason they are called symmetric is that they correspond to  a generalization of the polar decomposition of a linear matrix into a product of the orthogonal and a symmetric matrix
\be
\cV= e^{\phi\cdot {\bf \Sigma}} e^{\theta\cdot {\bf \Lambda}}
\label{polar}\ee
Here ${\bf \Lambda}$ are the generators of the H group and ${\bf \Sigma}$ are the coset generators. A symmetric gauge is a choice
\be
\theta=0
\label{symgauge}\ee

In symmetric gauge
\be
{\cal V}_{sym} (\phi^r) = e^{\phi^r K_r} \in \exp (\mathfrak{k})\, \qquad r=1,\dots,  n_{sc}
\label{sym}\ee
where $\{K_r\}$  is a basis of the coset algebra
$\mathfrak{k}$ defined in \rf{GK}.  Note that the coset representative in the symmetric gauge is not in a subalgebra  of G. {\it All scalars in the symmetric H-gauge occur in the gauge-fixed supergravity action non-polynomially}.

In  symmetric  gauge, the scalars $\phi^r$ transform in a linear representation of the maximal compact subgroup $H \subset G$, and global H-invariance of the Lagrangian is manifest. The 1-loop anomalies of global H-symmetry in this gauge were computed in  \cite{Marcus:1985yy}.

\item {\it  Iwasawa triangular   gauges}

The early class of gauges relating D-dimensional supergravity to 11D supergravity was proposed and studied in \cite{Cremmer:1979up,Cremmer:1997ct}.

Iwasawa triangular gauges/coset space representatives  discussed in general in \cite{Samtleben:2008pe,Trigiante:2016mnt} are associated with  the Iwasawa decomposition of G with respect to  H
and with a solvable parametrization so that ${\cal V}_{Iwasawa} (\varphi^r)$  belongs to a solvable Lie group $G_S= exp (\mathscr {S})$
\be
\cV_{Iwasawa} (\varphi^r)= e^{\varphi^r T_r}\in \exp(\mathscr {S})\,,\label{solpar}
\ee
Here  $\{T_r \}$ is a basis of $\mathscr {S}$ $(r = 1,..., n_{sc})$, it is also known as a Borel subalgebra of $\mathfrak {g}$. The algebra $\mathscr {S}$, parametrized by the scalar fields of the theory,  has the following general structure:
\begin{equation}
\mathscr {S}={\tt C}\oplus {\tt N}\,,\label{SCN}
\end{equation}
where ${\tt C}$ is the Cartan subspace of the coset space $\mathfrak{k}$ and is defined as the maximal set of commuting semisimple generators, and ${\tt N}$ is a nilpotent  subalgebra. When the theory originates from dimensional reduction of a higher dimensional one, ${\tt C}$ is parametrized by the \emph{dilatonic moduli}.
The  space ${\tt N}$,  is parametrized by \emph{axionic fields}.
Since ${\tt N}$ is nilpotent, {\it the scalars parametrizing ${\tt N}$ appear in the supergravity action
polynomially, whereas the  dilatonic scalars parametrizing the  Cartan subalgebra ${\tt C}$ occur non-polynomially, in particular exponentially.}

Iwasawa gauges in supergravity include  cases where D-dimensional supergravity is derived from  maximal 11D, they are described by parabolic subgroups  $P_{\alpha_{2}}$ in Fig. \ref{dynkin}.
These
triangular gauges are discussed in \cite{Cremmer:1979up,Samtleben:2008pe} and presented in details  with  examples in \cite{Cremmer:1997ct}.
We discuss these triangular Iwasawa gauges in Appendix \ref{App:11}.

  \item {\it Partial Iwasawa gauge}

These gauges arise from D+1 dimensional supergravity where part of the scalars are
 in $G_{D+1}/H_{D+1}$. The freedom to gauge-fix a local $H_{D+1}$ symmetry includes cases where we can use
  a symmetric gauge for fixing local $H_{D+1}$ symmetry. This choice will produce a partial Iwasawa  gauge. In addition to known examples in  4D related to 5D in \cite{Sezgin:1981ac,Andrianopoli:2002mf},  we will show analogous examples of partial Iwasawa gauges in 6D related to 7D. In supergravity II in \cite{Cowdall:1998rs} the action still has a local $SO(5)$ symmetry associated with  a 7D  $({G/ H})_7={SL(5)\over SO(5)}$. We will gauge-fix the corresponding action of 6D supergravity II  in Iwasawa as well as in partial Iwasawa  gauges.

\end{itemize}
The first issue with regard to these choices of gauges is related to 1-loop anomalies, computed in symmetric gauges in  \cite{Marcus:1985yy}. For example, in maximal 4D supergravity the global H=SU(8) anomalies cancel and in 6D the global H=SO(5)x SO(5) anomalies cancel.
However, supergravities in Iwasawa type gauges  do not have manifest global $H_D$ symmetries, rather they have the properties related to theories in higher dimensions. For example, the parabolic groups $P_{D}$ are shown in Fig. \ref{EA}.  In 4D  $P_{7}$ subgroup of $E_7$   has the node 7 removed and we get $E_{6}$, a symmetry of D+1=5 supergravity. In 6D  $P_{5}$ subgroup of $E_5$  has the node 5 removed and we get $A_{4}$, a symmetry of D+1=7 supergravity.

Therefore the relation between the quantum field theories  in these different gauges, symmetric and Iwasawa, requires an investigation.

The gauge-fixing of maximal 4D supergravity is known in a symmetric gauge \cite{Cremmer:1979up,deWit:1982bul}, and in
$\cN\geq 5$ in 4D with  ${G\over H}={SO^*(12)\over  U(6)}$ and ${G\over H}= {SU(1,5)\over  U(5)}$ for $\cN=6,5$ respectively, by truncation, or
 by using  symplectic sections as in \cite{Andrianopoli:1996ve}. But less is known about Iwasawa type  gauges, and no relation between these theories in different gauges was ever  established. In 6D maximal supergravity no gauge-fixing of a local $SO(5)\times SO(5)$ was performed, neither in symmetric nor in a triangular gauge.

There is a certain relation between  versions of supergravity constructed in various dimensions D with G/H coset space, and supergravities obtained by Scherk-Schwarz generalized dimensional reduction from higher D \cite{Scherk:1979zr}. It  was explained in \cite{Cremmer:1979up,Cremmer:1997ct}, it all boils down to gauge-fixing local H symmetry.  The choice of the gauge, suggested in  \cite{Cremmer:1979up}  is ``up to the user'': one can have a symmetric gauge \cite{Cremmer:1979up,deWit:1982bul} where the asymptotic fields transform in the linear representation of H and dependence on all scalars is non-polynomial. Or one can have  Iwasawa  or partial Iwasawa gauges, where axionic scalars enter only polynomially.
However, the local H-symmetry must be anomaly-free for the S-matrix to be independent of the user's choice!
\section{Gauge-fixing  4D maximal supergravity I}\label{Sec:4D}
\subsection{Symmetric gauge}

We start with  a standard (ungauged) Lagrangian in 4D \cite{Cremmer:1979up, deWit:1982bul} with  a manifest  $SL(8, \mathbb{R})$ symmetry which is a subgroup of \E\,.  The duality representation ${\bf 56}$ of ${\rm E}_{7(7)}$ and its adjoint representation ${\bf 133}$ branch with respect to ${\rm SL}(8,\mathbb{R})$ as follows:
    \begin{align}
    {\bf 56}&\rightarrow {\bf 28}+{\bf 28}'\,,\nonumber \\
    {\bf 133}&\rightarrow {\bf 63}+{\bf 70}\,.\label{sl8branch}
    \end{align}
A general polar decomposition presented in eq. \rf{polar} and the corresponding choice of the symmetric gauge \rf{symgauge} in case of ${G\over H}= {E_{7(7)}\over SU(8)}$ is specified as follows.
 To perform gauge-fixing      requires to put a restriction on the 56-bein ${\cal V}$ depending on 133 scalars, so that it will depend only on 70 scalars, coset space coordinates. The restriction is for the 56-bein  to take a form ${\cal V}_{gf}= {\cal V}_{gf} ^{\dagger}$ \cite{Cremmer:1979up, deWit:1982bul} \be
{\cal V} \qquad \to  \qquad {\cal V}_{gf} = {\cal V}_{gf} ^{\dagger} = e^{X}= \begin{pmatrix}\cosh \phi \bar \phi& \phi \frac{\sinh \bar \phi \phi}{\bar \phi \phi}\cr\bar  \phi \frac{\sinh \phi \bar\phi}{ \phi \bar\phi}&\cosh \bar \phi \phi\end{pmatrix}\,,  \qquad  X= \left(\begin{array}{cc}0 & \phi_{ijkl} \\\bar \phi^{mnpq} & 0\end{array}\right)  \,.
\ee
Here the self-dual scalars $\phi_{ijkl}= \pm {1\over 4!} \epsilon_{ijklpqmn}\bar \phi^{pqmn}$ transform in the 35-dimensional representation of $SU(8)$, and is considered as a $28\times 28$ matrix in the antisymmetric combinations $[ij],\,[k\ell]$.  Indices are raised by complex conjugation. $\bar \phi \phi $ stands for the matrix $(\bar \phi \phi)^{ij}{}_{k\ell}= \frac{1}{2}\bar \phi^{ijmn}\phi _{mnk\ell}$. It is useful to use the inhomogeneous coordinates of \E/SU(8)
\be
y_{ij, kl}= \Big (\phi {\tanh\sqrt{(\bar \phi \phi)}\over \sqrt{(\bar \phi \phi)}}\Big )_{ij, kl}= \frac{1}{2}\phi _{ijmn}\left({\tanh\sqrt{(\bar \phi \phi)}\over \sqrt{(\bar \phi \phi)}}\right)^{mn}{}_{k\ell}\,,
\ee
such that
\begin{equation}
  e^X =  \left(\begin{array}{cc}{1\over \sqrt{1-y \bar y}} & y {1\over \sqrt{1-\bar y y}}\\
\, \, \bar y \, {1\over \sqrt{1-y \bar y}}  &\, \,  {1\over \sqrt{1-\bar y y}} \end{array}\right)\,.
 \label{eXiny}
\end{equation}
In this way the 63 local parameters of $SU(8)$ are used to define the  unitary  symmetric gauge with 70 physical scalars, in agreement with eq.  \rf{sym} above.
Details showing that after gauge-fixing  local $H= $SU(8) symmetry, there is a remaining nonlinearly realized \E\, and   a  field-dependent compensating H= SU(8)  symmetry, are given in \cite{Cremmer:1979up, deWit:1982bul,Kallosh:2008ic,Bossard:2010dq}.
 The scalars $\phi_{ijkl}, \, \bar \phi^{pqmn}$ transform in a linear representation of $SU(8)$. The 1-loop anomalies of global $SU(8)$-symmetry in this gauge cancel \cite{Marcus:1985yy}.

\subsection{Iwasawa D+1 gauge}\label{Sec:4Dparab}

In terms of Fig. \ref{dynkin} this is the parabolic group $P_{\alpha_7}$ when the $\alpha_7$ node is removed. In Fig. \ref{EA} at the left panel we also see the Dynkin diagram of $E_7$ and the one for $E_6$, once the right node was removed.

In attempt to understand the 5D/4D connection we can make a choice of the Iwasawa  gauge for the  action in \cite{Cremmer:1979up, deWit:1982bul}  based on the fact that in 5D scalars are in ${E_{6(6)}\over USp(8)}$ coset space.  With respect to ${\rm E}_{6(6)}\times {\rm SO}(1,1)$
the ${\bf 56}$ and the adjoint of ${\rm E}_{7(7)}$ decompose instead of \rf{sl8branch} as follows:
\begin{eqnarray}
{\bf 56}&\rightarrow&   {\bf
1}_{-3}+{\bf 27}'_{-1}+{\bf 1}_{+3} +    {\bf 27}_{+1} \,,
\nonumber\\
{\bf 133}&\rightarrow&{\bf 27}_{-2}+ {\bf 1}_0+{\bf 78}_0+{\bf 27}'_{+2}\,.\label{133e6}
\end{eqnarray}
Here {\bf 28} vectors split into ${\bf
1}_{-3}$ from a 5D metric and  27 vector fields of 5D in ${\bf 27}'_{-1}$ of the electric group. The ${\bf 78}_0$ scalars are related to generators of $E_{6(6)}$, it is a part of \E\, generators. The ${\bf 27}'_{+2}$ scalars are  27 axionic scalars originating from the five-dimensional vector fields through the Kaluza-Klein reduction. The  ${\bf 1}_0$ is related to a radius of the 5th dimension. We can now choose a gauge-fixing of a 56-bein $\cV$ in the action of standard 4D supergravity in \cite{Cremmer:1979up, deWit:1982bul} in the form depending on 70 scalars associated with 5D maximal supergravity.

The  Lagrangian \cite{Cremmer:1979up, deWit:1982bul} has 63 local H-symmetries. In Iwasawa gauge 36 of these can be  associated with a local H-symmetry of 5D supergravity, $USp(8)$, but there are 27 more which we can use to make a gauge-fixing related to compactified 5D supergravity.
\be
{\bf 63}= {\bf 36}+ {\bf 27}
\ee
Our 70=1+42+27 scalars in notation of \cite{Trigiante:2016mnt} consist of
\be
1+42+27:  \qquad \sigma, \, \hat \phi, \, a^\l
\ee
Here in ${\bf 133}\, \rightarrow \, {\bf 27}_{-2}+ {\bf 1}_0+{\bf 78}_0+{\bf 27}'_{+2}$ we are using 36 local symmetries to get  78-36=42 of $\hat \phi$, and we use 27 local symmetries to remove ${\bf 27}_{-2}$ scalars. Thus there are 2 possible  gauge-fixing condition on the 56-bein. The first one is the triangular
  Iwasawa gauge
\be
\cV_{gf}= e^{a^\l t_\l} \, e^{\hat \phi^r \hat T_r} \, e^{\sigma D} \qquad r=1,\dots , 42
\label{Mario}\ee
Here the expressions for the 56x56 matrices, operators $D, t_\l$ are given in eq. (4.26) in \cite{Trigiante:2016mnt} and 27 $t_\l$ form an algebra $[t_\l, t_\delta] =0$. The coset space ${E_{6(6)}\over USp(8)}$ representative for the 42 scalars in $e^{\hat \phi^r \hat T_r}$ is in the solvable parametrization,  with solvable Lie algebra $\mathscr{S}$.

The choice (\ref{Mario}) corresponds to describing the scalar manifold of the theory as isometric to the following manifold  \cite{Trigiante:2016mnt} \begin{equation}
{\cal M}_{scal} \sim \left[{\rm O}(1,1)\times\frac{{\rm E}_{6(6)}}{{\rm USp}(8)} \right]\ltimes \exp({\tt N}^{[{\bf 27}'_{+2}]})\,,
\end{equation}
where $\ltimes$ denotes the semi-direct product and ${\tt N}^{[{\bf 27}'_{+2}]}$ is the 27-dimensional space spanned by $t_\lambda$. If we choose for ${E_{6(6)}\over USp(8)}$ coset  the solvable parametrization, then (\ref{Mario}) defines the solvable parametrization of ${\cal M}_{scal}$, with solvable Lie algebra ${\Scr S}$. Using the property that the $t_\lambda$ are commuting, one can verify that the scalars $a^\lambda$ are covered by a derivative and
the constant shifts $a^\lambda\rightarrow a^\lambda+c^\lambda$ are isometries. They are implemented by the ${\rm E}_{7(7)}$-transformation ${\bf g}=e^{c^\lambda\,t_\lambda}$:
\begin{equation}
{\bf g}\,\cV_{gf}(a^\Lambda,\hat{\phi},\sigma)=e^{(a^\l + c^\l) t_\l} \, e^{\hat \phi^r \hat T_r} \, e^{\sigma D}\,.
\end{equation}

\subsection{Partial Iwasawa D+1 gauge}\label{Sec:4Dpartial}
\be
\cV_{gf}= e^{a^\l t_\l} \, e^{\phi^{abcd} K_{abcd}} \, e^{\sigma D}
\ee
Here the expressions for the 56x56 matrices, operators $D, t_\l$ are given in eq. (4.26) in \cite{Trigiante:2016mnt} and 27 $t_\l$ form an algebra $[t_\l, t_\delta] =0$. But the  coset representative for the 42 scalars $\phi^{abcd} $ in
${E_{6(6)}\over USp(8)}$ coset space is not taken in  the solvable parametrization. Instead it is taken in a symmetric gauge in the form   $e^{\phi^{abcd} K_{abcd}}$.  It is the one which was chosen in \cite{Andrianopoli:2002mf},  it correspond to a symmetric gauge in 5D in  \cite{Sezgin:1981ac} and  leads to a partial Iwasawa gauge in 4D in \cite{Andrianopoli:2002mf}.

\section{ Supergravity I and   supergravity II in 4D }\label{Sec:5D4D}

The maximal supergravity I  in 4D \cite{Cremmer:1979up, deWit:1982bul} is rather well known, whereas supergravity II \cite{Andrianopoli:2002mf} is not well known. It was developed and applied  in \cite{Ceresole:2009jc} in the context of extremal black holes and $\cN=8$ attractors \cite{Ferrara:2006em}.
The non-BPS regular extremal black holes turned out to be solutions of  supergravity II with its $E_{6(6)}$ basis.  The 1/8 BPS extremal black holes  entropy  is given by a quartic  \E\, invariant, which is also manifestly SU(8) invariant. But the non-BPS black hole with the regular horizon has  the entropy which is manifestly $E_{6(6)}$ invariant,   it depends on a cubic $E_{6(6)}$ invariant. These black holes are in different  orbits of the fundamental representations of the exceptional groups \E\,  and $E_{6(6)}$
\cite{Ferrara:1997ci,Ferrara:1997uz}.

The structure of
 maximal supergravity II \cite{Andrianopoli:2002mf}  in 4D was  influenced by  5D supergravity, but it was constructed directly in 4D by decomposing \E\, symmetry
\begin{equation}\fe_{7,7}= \fe_{6,6}+\fso(1,1)+ \fp,\qquad
\fp=\mathbf{27_{-2}}+\mathbf{27'_{+2}},\label{e77}\end{equation}
where $\fp$ carries the  representations $\mathbf{27_{-2}}$ and $\mathbf{27'_{+2}}$
of
$\fe_{6,6}+\fso(1,1)$. It was also shown to agree with the one
derived from 5D by compactification in \cite{Sezgin:1981ac} in the limit of vanishing gaugings.  An ungauged version of it was presented in \cite{Andrianopoli:2002mf} when all mass parameters/gauge couplings vanish.

The ungauged Lagrangian of  supergravity in 4D in  \cite{Andrianopoli:2002mf}, which we call supergravity II, does not have local symmetries since the reduction in \cite{Sezgin:1981ac} started in 5D supergravity with ${G\over H}=
{E_{6(6)}\over USp(8)}$ where the 5D local USp(8) symmetry is gauge fixed in a symmetric gauge. It means that from 78 scalars of $E_{6(6)}$  the 36 are eliminated using the local USp(8) symmetry, and only 42 are left. It also means that $E_{6(6)}$ symmetry is realized nonlinearly.

In addition to 42 scalars $\phi^{abcd}$, coset coordinates of ${E_{6(6)}\over USp(8)}$, the extra 28 scalars in 4D come from the size of the extra dimension circle $\phi$ and from  the extra component of the 27-dimensional vector in 5D $a^\Lambda = A_5^\Lambda$. Therefore the 70 scalars in  \cite{Andrianopoli:2002mf}  are
\be
 \phi\, , \, a^\Lambda\, , \,  \phi^{abcd}
\ee
It means that ${\bf 70}$ of SU(8) are decomposed as follows under USp(8)
\be
{\bf 70}\to {\bf 1}+{\bf 27}+{\bf 42}
\ee
This is to  be compared with 133 scalars in standard 4D maximal supergravity \cite{Cremmer:1979up, deWit:1982bul}
\be
 {\bf 133} \rightarrow {\bf 63}+{\bf 70}
\ee
or in  ${\rm E}_{6(6)}\times {\rm SO}(1,1)$ decomposition
\be
{\bf 133} \rightarrow {\bf 27}_{-2}+ {\bf 1}_0+{\bf 78}_0+{\bf 27}'_{+2}
\ee
in \cite{Andrianopoli:2002mf}. Now we need to consider gauge-fixing local SU(8) in \cite{Cremmer:1979up, deWit:1982bul} in a way to reproduce either 70 scalars transforming linearly under global SU(8), or getting 70  by removing ${\bf 27}_{-2}$ scalars as well as 36 out of ${\bf 78}_0$. The first case is a standard unitary gauge called symmetric gauge in \cite{Cremmer:1979up, deWit:1982bul}. The second case related to   supergravity II in   \cite{Andrianopoli:2002mf} is the choice we called a partial Iwasawa gauge.

It might be useful to bring up the scalar action in \cite{Andrianopoli:2002mf,Ceresole:2009jc} for a better understanding of the issues of gauge-fixing local symmetries in 4D supergravities, in particular the difference between symmetric and Iwasawa gauges. The scalar action in \cite{Andrianopoli:2002mf,Ceresole:2009jc} is
\be
{1\over e}\,\cL^{sc}_{_{5D4D}}= {3\over 2} \partial_\mu \phi \partial^\mu \phi -{1\over 4} e^{-4\phi} \hat {\cal N} _{\Lambda \Sigma}  \partial_\mu a^\Lambda  \partial^\mu a^\Sigma +{1\over 4!} P_\mu{}^{abcd} P^\mu{}_{abcd}
\ee
Here $\hat {\cal N} _{\Lambda \Sigma}$ is the 5D ($SO(1,1)$ invariant) vector kinetic matrix, $\Lambda=1,\dots 27$,  $a=1,\dots , 8$ and 70 scalars are decomposed under USp(8) as ${\bf 70}\to {\bf 1}+{\bf 27}+{\bf 42}$.

In the supergravity action \cite{Cremmer:1979up, deWit:1982bul} the pure scalar part, before gauge-fixing local SU(8) symmetry has the form
\be
{1\over e} \cL^{sc}_{_{ 4D}} = {1\over 4!} P_\mu{}^{ijkl} P^\mu{}_{ijkl}
\ee
where $
({\cV}^{-1} D_\mu \cV)^{ijkl} = {\cal P}_\mu^{ijkl}
$ is a local SU(8) tensor in ${\bf 70}$.

 The 28 vectors in the action in \cite{Cremmer:1979up, deWit:1982bul} in \cite{Andrianopoli:2002mf} are represented by 1+27 vectors in  \cite{Andrianopoli:2002mf,Ceresole:2009jc}
 \be
B_\mu\, , \, Z_\mu^\Lambda
\ee
 Both actions depend only on field strength's: 28 in  $F_{\mu\nu}^{IJ} =\partial_\mu \cA_\nu^{IJ} -\partial_\nu \cA_\mu^{IJ} \,$ in \cite{Cremmer:1979up, deWit:1982bul} and 1+27 $B_{\mu\nu} =\partial_\mu B_\nu-\partial_\nu B_\mu\,$ and $Z_{\mu\nu}^\Lambda =\partial_\mu B_\nu^\Lambda-\partial_\nu B_\mu^\Lambda\,$ in  \cite{Andrianopoli:2002mf}.
 The scalar-vector Lagrangian is
\bea
&&\frac{1}{e}{\cal L}^{vec}=
 {\cal I}_{00}(\phi)\,B_{\mu\nu} B^{\rho\sigma}+ 2 {\cal I}_{0\Lambda}(\phi)\,B_{\mu\nu}\,Z^{\Lambda\,\mu\nu}+{\cal I}_{\Lambda\Sigma}(\phi)\,Z^\Lambda_{\mu\nu}\,Z^{\Sigma\,\mu\nu}
\cr
\cr
&+&\frac{1}{2\,e} \epsilon^{\mu\nu\rho\sigma}  [ {\cal R}_{00}(\phi) B_{\mu\nu} B_{\rho\sigma}  +2{\cal R}_{0\Lambda}(\phi) B_{\mu\nu} \,Z^{\Lambda}_{\rho\sigma}+{\cal R}_{\Lambda \Sigma}(\phi) Z^\Lambda_{\mu\nu} \,Z^{\Sigma}_{\rho\sigma}]
\label{boslagr}
\eea
where ${\cal I}_{IJ}$ and ${\cal R}_{IJ}$ are given by 4 blocks with 28 split into 0 and 27 $\Lambda$'s
\ba\label{im-enne}  {\cal I}_{IJ} &=& \left(
\begin{array}{c|c}
&\\
{\cal I}_{00}\ &\ {\cal I}_{0\, \Sigma}\\&\\
\hline\\ \vspace{-1pt}
{\cal I}_{\Lambda\,0}\ & {\cal I}_{\Lambda\, \Sigma}\\&\\
\end{array}\right)=
-
e^{6\phi}\left(
\begin{array}{c|c}
&\\
1+ e^{-4\phi}a_{\Lambda \Sigma}\,a^{\Lambda}a^{\Sigma}
\ &\ -e^{-4\phi}{a}_{\Lambda \Sigma}\,a^{\Lambda}
\\&\\
\hline\\ \vspace{-1pt}
-e^{-4\phi}a_{\Lambda \Sigma}\,a^{\Sigma}\ &
e^{-4\phi} a_{\Lambda \Sigma}\\&\\
\end{array}
\right)
; \ea
and
 \ba\label{re-enne} {\cal R}_{IJ}&=& \left(
\begin{array}{c|c}
&\\
{\cal R}_{00}\ &\ {\cal R}_{0\, \Sigma}\\&\\
\hline\\ \vspace{-1pt}
{\cal R}_{\Lambda\,0}\ & {\cal R}_{\Lambda\, \Sigma}\\&\\
\end{array}\right)=
 \left(
\begin{array}{c|c}
&\\
\frac13d_{\Lambda \Sigma \Gamma}a^{\Lambda}a^{\Sigma}a^{\Gamma} \ &\
-\frac12d_{\Sigma \Lambda \Gamma}a^{\Lambda}a^{\Gamma}\\&\\
\hline\\ \vspace{-1pt}
-\frac12d_{\Lambda \Sigma \Gamma}a^{\Sigma}a^{\Gamma}\ &
\  d_{\Lambda \Sigma \Gamma}a^{\Gamma}\\&\\
\end{array}\right)\ ,\ea
Here $d_{\Lambda \Sigma \Gamma}$
 is the symmetric invariant tensor of the representation {\bf 27} of $E_{6(6)}$ and $a_{\Lambda \Sigma}$ is
 the five dimensional
(SO(1, 1) invariant) vector kinetic matrix. Note that
 the scalar-dependent kinetic terms of vectors
 are polynomial in axions $a^{\Lambda}$.

 The scalar-vector Lagrangian in a symmetric gauge in 4D is
 \begin{equation}
\frac{1}{e}{\cal L}^{vec}=
\frac{1}{4}\, {\cal I}_{IJ}(\phi)\,F^I_{\mu\nu}\,F^{J\,\mu\nu}
+\frac{1}{8\,e}\,{\cal R}_{IJ}(\phi)\,\epsilon^{\mu\nu\rho\sigma}\,F^I_{\mu\nu} \,F^{J}_{\rho\sigma}\,,
\label{boslagr2}
\end{equation}
 Here $I,J =1,\dots , 28$.
 See for example, eq. (2.1) in \cite{Trigiante:2016mnt}.

 Both actions 4D, supergravity I and II,  when  supplemented with fermions have maximal local supersymmetry. Supergravity II has  inherited local supersymmetry via dimensional reduction. The gauged supergravity action in  \cite{Sezgin:1981ac} has spontaneously broken supersymmetry, however, it is restored in the ungauged supergravity with  $m_i\to 0$.

\section{Gauge-fixing  6D maximal supergravity I}\label{Sec:6D}

Explicit gauge-fixing  of 6D maximal supergravity  is interesting, in particular, since Marcus anomaly cancellation is valid for a symmetric gauge in 6D  \cite{Marcus:1985yy}.  But
 the gauge-fixing of local $ SO(5) \times SO(5)$ symmetry in 6D supergravity was not done before.

There is a technical complications due to presence in the action of two  kinds of vielbeins, 10x10 and 16x16 \cite{Tanii:1984zk,Bergshoeff:2007ef} vielbeins, related to each other. It is therefore necessary to find coset representatives for both of these, and check that the relations between them are valid upon gauge-fixing.
This is  different from 4D where there is only 56x56  vielbein \cite{Cremmer:1979up, deWit:1982bul}.

We consider 6D supergravity \cite{Tanii:1984zk,Bergshoeff:2007ef}
and try to gauge-fix local $ SO(5) \times SO(5)$ symmetry. Before gauge-fixing scalars form an $SO(5, 5)$ valued 16x16 vielbein matrix $V_{\mu\dot \mu}{}^{\alpha \dot \alpha}$ with 45 independent entries $\mu, \dot \mu =1,2,3,4, \, \alpha, \dot \alpha =1,2,3,4$. When two local $ SO(5) \times SO(5)$ symmetries are gauged-fixed, the 25 scalars which are left are representative of the coset space ${\cG\over \cH}={E_{5(5)}\over USp(4)\times USp(4)} \sim {SO(5, 5)\over SO(5) \times SO(5)}$.

The theory has a symmetry in which
 the first $SO(5) $ is flipped into the second $SO(5)$  and chirality flips. For example, spin 1/2 fields are singlets of $E_{5(5)}$  and are either in (5,4) or (4,5) of $ SO(5) \times SO(5)$, depending on their chirality.

 \subsection{Symmetric  gauge}

Now we consider 6D supergravity \cite{Tanii:1984zk,Bergshoeff:2007ef}
and try to gauge-fix local $ SO(5) \times SO(5)$ symmetry.\footnote{We have repeated and extended their notations in Appendix \ref{app:gammas}.}
The $10\times 10$ coset  representative of ${SO(5, 5)\over SO(5) \times SO(5)}$ is available in explicit form in \cite{Gilmore:2008zz}.

\

{ \it $10\times 10$   representation of the vielbein}

\

\noindent The $10 \times 10 $ $SO(5, 5)$ matrix $U$ in eq. (8) in  \cite{Tanii:1984zk} satisfies the constraint
\be
U^T\eta_d \, U=\eta_d  \qquad \eta_d= \left(\begin{array}{cc}I & 0 \\0 & -I\end{array}\right)
\label{5.5}
\ee
For
\be
U= \left(\begin{array}{cc}A & B \\C & D\end{array}\right)
\label{U}\ee
eq. \rf{5.5} means that
\be
A^T A -C^T C =1\, , \quad B^T B- D^T D=-1\, , \quad A^TB - C^T D= B^TA-D^TC=0
\label{U1}\ee
Under $SO(5, 5)$ matrix $g$ and $ SO(5) \times SO(5)$ matrix $h(x)$ it transforms as
\be
U'= g\, U \, h^{-1} (x)
\ee
It has  45 independent entries and $g$ and $h$ are linearly realized and independent before local $ SO(5) \times SO(5)$ is gauge-fixed. We can  realize this unitary gauge where only 25 scalars are left, after local 10+10 parameters of $ SO(5) \times SO(5)$ are used. Namely, we impose the  condition on $U$ to gauge-fix local $ SO(5) \times SO(5)$ symmetry in the form\footnote{Note also that in 4D the related choice of a symmetric gauge in  \cite{Cremmer:1979up, deWit:1982bul}
 is for the $56\times 56$-bein ${\cal V}_{gf}= {\cal V}_{gf}^{\dagger}$}
\be
U_{g.f.}= U^T_{g.f.}
\ee
in agreement with
 the coset space construction for ${SO(r, s)\over SO(r) \times SO(s)}$ in \cite{Gilmore:2008zz}.  In such case we get
\be
U_{g.f.}=\exp
\left(\begin{array}{cc}0 & \phi \\\phi^T & 0\end{array}\right) =
\left(\begin{array}{cc}\cosh\sqrt {(\phi \phi^T)} & \phi \, {\sinh \sqrt {(\phi^T \phi)} \over \sqrt {(\phi^T \phi)} } \\ \, \, \phi^T \, {\sinh \sqrt {(\phi \phi^T)} \over \sqrt {(\phi \phi^T)} }  &\, \,  \cosh\sqrt {(\phi^T \phi)} \end{array}\right)
\label{gf}\ee
Here the asymptotic field $\phi_{a\dot a}$ is in (5,5) of $ SO(5) \times SO(5)$, $a, \dot a=1, \dots , 5$.
Note that the matrices $\phi^T \phi$ and $\phi \phi^T$ (with $\phi ^T$ written as $(\phi ^T)^{\dot aa}= \delta ^{\dot a\dot b}\phi _{b\dot b}\delta ^{ba}$) are hermitian with nonnegative values and have  nonzero eigenvalues identical. The square roots do not appear in the final expressions, since the entries in (\ref{gf}) are power series with even powers of $\sqrt{\phi \phi^T}$.

As in \cite{Cremmer:1979up, deWit:1982bul,Kallosh:2008ic,Bossard:2010dq} where the 4D supergravity local $SU(8)$ symmetry was gauge-fixed, we introduce a set of inhomogeneous coset coordinates
\begin{align}
&y_a{}^{ \dot a} = \phi_a{}^{\dot b} \Big ({\tanh \sqrt {(\phi^T \phi)} \over \sqrt {(\phi^T \phi)} }\Big )_{\dot b}{}^{ \dot a}=\Big ({\tanh \sqrt {(\phi\phi^T)} \over \sqrt {(\phi\phi^T )} }\Big )_{a}{}^{b} \phi _b{}^{\dot a}\,,\qquad (y^T)_{\dot a}{}^a = \delta _{\dot a\dot b}\phi _b{}^{\dot b}\delta ^{ba}\,,
\nonumber\\
&(U_{g.f.}^{sym})_{\underline{A}}{}^{\underline{B}} =\left(\begin{array}{cc}(U_{g.f.})_a{}^b&(U_{g.f.})_a{}^{\dot b}\\ \\ (U_{g.f.})_{\dot a}{}^b&(U_{g.f.})_{\dot a}{}^{\dot b}\end{array}\right)= \left(\begin{array}{cc}{1\over \sqrt{1-y y^T}} & y {1\over \sqrt{1-y^T y}}  \\
\\
\, \, y^T \, {1\over \sqrt{1-y y^T}}  &\, \,  {1\over \sqrt{1-y^T y}} \end{array}\right)
\label{gf1alt}
\end{align}

\

{\it $16\times 16$   representation of the vielbein}

\

\noindent The 45 generators of $SO(5, 5)$  in spinorial representation can be given in $32\times 32$ form, and then split in chiral and antichiral $16\times 16$ matrices.
\begin{align}
&\SO(5,5) \mbox{ representation }\frac{1}{2}\Gammabig_{\underline{A}\underline{B}}= \frac{1}{2}\begin{pmatrix} \Gamma _{\underline{A}\underline{B}} & 0 \cr 0& \Gamma' _{\underline{A}\underline{B}}\end{pmatrix}\,,\nonumber\\
&10: \tilde \Lambda_{[ab]}= \Gammabig_{ab}\,,\qquad \Lambda_{[ab]}=\Gamma_{ab}=\Gamma '_{ab}= \unity _4\times \gamma_{ab}\,,\nonumber\\
& 10: \tilde \Lambda_{[\dot a\dot b]}= \Gammabig_{\dot a\dot b}\,,\qquad \Lambda_{[\dot a\dot b]}=\Gamma _{\dot a\dot b}=\Gamma' _{\dot a\dot b}= -\gamma_{\dot a\dot b}\times\unity _4\,,\nonumber\\
&15: \tilde \Sigma_{a\dot a}= \Gammabig_{a\dot a}\,,\qquad  \Sigma_{a\dot a}=\Gamma     _{a\dot a}=- \Gamma '    _{a\dot a}= -\gamma_{\dot a}\times\gamma _a\,.
   \label{SO55reps}
\end{align}

\noindent Before gauge-fixing there are  $32\times 32$ matrices, block diagonal
\begin{align}
W=& \exp\left[{\frac{1}{4}\phi^{\underline{A}\underline{B}}(\Gammabig_{\underline{A}\underline{B}})_{{\cal B}}{}^{{\cal A}}}\right]\nonumber\\
&\exp \left[{\frac{1}{2}(\phi_a{}^{\dot a} \tilde \Sigma^{a}{}_{\dot a})}+\frac{1}{4}(\theta_{ab} \tilde \Lambda^{[ab]} + \frac{1}{4}\theta_{\dot a\dot b} \tilde \Lambda^{[\dot a\dot b]})\right] \,.
\label{Wbeforegf}
\end{align}

We can split $W$ into 16x16 matrices,  chiral and antichiral parts
\begin{align}
 W_{{\cal A}}{}^{{\cal B}} =\begin{pmatrix}\bar V{}_A{}^B&0\cr 0&V_A{}^B\end{pmatrix}\,, \qquad \mbox{with }\qquad {\bar V}^B{}_A=C^{BC}\bar V_C{}^DC_{DA}= (V^{-1})_{A}{}^B\,.
   \label{WinVbarV}
\end{align}
Thus $\bar V$ is  the transpose of the inverse of $V$, i.e.
\begin{equation}
  V_A{}^C \bar V^B{}_C = \delta _A^B\,,\qquad V_A{}^C\bar V^A{}_D=\delta ^C_D\,.
 \label{VbarV}
\end{equation}
These matrices are related to $U_{\underline{A}}{}^{\underline{B}}$ in (\ref{U}), (\ref{U1}) by
the requirement of invariance of $(\Gamma_{\underline{A}})^{AB}$ and $(\Gamma'_{\underline{A}})_{AB}$ (the left-right and right-left chiral parts of $\Gammabig_{\underline{A}}^{AB} $) under simultaneous $\SO(5,5)$ action on all their indices:  \cite{Tanii:1984zk,Bergshoeff:2007ef} \footnote{Most equations in these references are written in a covariant basis,
useful to write down the action (for the split electric/magnetic fields). For the group theory, the $\underline{A}$ basis is more transparent. The equations are equivalent after multiplication with the matrix $M$ in \cite[(2.29)]{Bergshoeff:2007ef}.}
\begin{align}
 V_{A}{}^C(\Gamma _{\underline{B}})^{AB}V_B{}^D = &U_{\underline{B}}{}^{\underline{A}}(\Gamma _{\underline{A}})_{CD}\,,\nonumber\\
  \bar V^A{}_C (\Gamma'_{\underline{B}} )_{AB}\bar V^{B}{}_{D} =&U_{\underline{B}}{}^{\underline{A}}(\Gamma' _{\underline{A}})_{CD}\,.
    \label{VGV}
\end{align}
or
\begin{align}
  U_{\underline{B}}{}^{\underline{A}}= & \frac{1}{16}V_{A}{}^{C}(\Gamma _{\underline{B}})^{AB}V_B{}^D (\Gamma'{}^{\underline{A}})_{DC}\,,\nonumber\\
  = &\frac{1}{16} \bar V^A{}_C (\Gamma'_{\underline{B}} )_{AB}\bar V^{B}{}_{D} (\Gamma ^{\underline{A}})^{DC}\,.
  \label{extractingU}
\end{align}
Note first that taking $V=\bar V=\unity _{16}$, one obtains $U=\unity _{10}$ (the value for $\phi =0$ in the equations in \cite{Gilmore:2008zz}), see  (\ref{gf}).

Secondly, these expressions lead to a matrix $U$ that is in $\SO(5,5)$, i.e.
\begin{equation}
  U_{\underline{A}}{}^{\underline{C}}U_{\underline{B}\underline{C}}= (\eta _d)_{\underline{A}\underline{B}}\,.
 \label{U2}
\end{equation}
In fact, inserting the first line of (\ref{extractingU}) in the second factor $U$ of  (\ref{U2}), we get
\begin{align}
 U_{\underline{A}}{}^{\underline{C}}U_{\underline{B}\underline{C}}=   & U_{\underline{A}}{}^{\underline{C}}\frac{1}{16}V_{A}{}^C(\Gamma _{\underline{B}})^{AB}V_B{}^D (\Gamma'{}_{\underline{C}})_{DC}   \nonumber\\
  = & \frac{1}{16}V_{A}{}^{C}(\Gamma _{\underline{B}})^{AB}V_B{}^D    \bar V^E{}_D (\Gamma'_{\underline{A}} )_{EF}\bar V^{F}_{C} \nonumber\\
  = & \frac{1}{16}(\Gamma _{\underline{B}})^{AB}(\Gamma'_{\underline{A}} )_{BA} \nonumber\\
   =&(\eta _d)_{\underline{A}\underline{B}}\,,
  \label{calculateUU}
\end{align}
where for the second line the second line of (\ref{VGV}) is used, for the third line  (\ref{VbarV}) and finally (\ref{Trace16}) with an index lowered.

The  16x16 matrices are
\be
\bar V= e^{\frac{1}{4}\phi^{\underline{A}\underline{B}}\Gamma_{\underline{A}\underline{B}}}\, ,\quad V= e^{\frac{1}{4}\phi^{\underline{A}\underline{B}}\Gamma'_{\underline{A}\underline{B}}}
\ee

 Gauge fixing using 10+10 local $SO(5) \times SO(5)$
 \be
\theta_{ab}=  \theta_{\dot a\dot b}=0
 \ee
 and thus gauge fixed $W$ is
\begin{align}
W_{g.f}=  \exp {\frac{1}{2}(\phi_a{}^{\dot a}  \tilde \Sigma^{a}{}_{\dot a})}\,,\qquad W_{{\cal B}}{}^{{\cal A}} = -(W^{-1})^{\cal A}{}_{\cal B}=-\mathbb{C}^{{\cal A}{\cal C}} (W^{-1})_{{\cal C}}{}^{{\cal D}}\mathbb{C}_{{\cal D}{\cal B}}\,.
\label{VgfD6}
\end{align}

The exponential satisfies the latter equation, since
\begin{equation}
  (\Gammabig_{\underline{A}\underline{B}})_{{\cal B}}{}^{{\cal A}}=-(\Gammabig_{\underline{A}\underline{B}})^{\cal A}{}_{\cal B}\,,
 \label{Gammabig2}
\end{equation}
(see (\ref{D5chargeconj})) and for $\underline{A}=a$ and $\underline{B}=\dot a$, and the minus sign in (\ref{VgfD6}) leads from $W$ to $W^{-1}$.

Gauge fixed $\bar V,  V$ are
\begin{align}
\bar V_{g.f}=  \exp \left(\frac{1}{2} {\phi_a{}^{\dot a}   \Sigma^{a}{}_{\dot a}}\right)\,,\qquad   V_{g.f}=\exp\left( {- \frac{1}{2}\phi_a{}^{\dot a}   \Sigma^{a}{}_{\dot a}}\right)
\label{Vgf}
\end{align}

Finally, one can show that the linear part in $\phi $ of  (\ref{VgfD6}) leads with  (\ref{extractingU}) to the linear part of (\ref{gf}). For that it is convenient to write the two equations (\ref{extractingU}) as (remembering that the $\Gamma _{\underline{A}}$ and $\Gamma ' _{\underline{A}}$ are symmetric)
\begin{align}
  U_{\underline{B}}{}^{\underline{A}}&=  \frac{1}{16}\Tr\left[ V\Gamma'{}^{\underline{A}}V^T \Gamma _{\underline{B}}\right]=  \frac{1}{16}\Tr\left[ V\Gamma'{}^{\underline{A}}\bar V^{-1} \Gamma
  _{\underline{B}}\right]\nonumber\\
  &=
 \frac{1}{16} \Tr \left[\bar V \Gamma ^{\underline{A}} \bar V^T \Gamma'_{\underline{B}}\right]=
 \frac{1}{16} \Tr \left[\bar V \Gamma ^{\underline{A}} V^{-1} \Gamma'_{\underline{B}}\right]\,,
 \label{VasTrace}
\end{align}
where we used that $V^T = \bar V^{-1}$.
These are traces over $16\times 16$ matrices. We can rewrite this expression  in $32\times 32$ language, which makes it easier to handle $\Gamma $ matrices.
Using  (\ref{VgfD6}), the equations (\ref{VasTrace}) are the block diagonal parts of
\begin{align}
  U_{\underline{B}}{}^{\underline{A}}=  &\frac{1}{32}\Tr \left[W\Gammabig ^{\underline{A}}W^{-1} \Gammabig _{\underline{B}}\right]\nonumber\\
  = & \frac{1}{32}\Tr \left[\begin{pmatrix}\bar V&0\cr 0&V\end{pmatrix} \begin{pmatrix}0&\Gamma^{\underline{A}}\cr \Gamma '{}^{\underline{A}}&0\end{pmatrix}
  \begin{pmatrix}\bar V^{-1}&0\cr 0&V^{-1}\end{pmatrix} \begin{pmatrix}0&\Gamma_{\underline{B}}\cr \Gamma '_{\underline{B}}&0\end{pmatrix}\right]\,.
  \label{UinW}
\end{align}

Now we can take standard $\gamma $-algebra, and use that traces of $\Gammabig _{\underline{A}\underline{B}}$ and $\Gammabig _{\underline{A}\underline{B}\underline{C}\underline{D}}$ vanish.
If we take the linear part of  (\ref{Wbeforegf})
\begin{equation}
 W =\unity _{32} = \frac{1}{4} \phi ^{\underline{A}\underline{B}}\Gammabig_{\underline{A}\underline{B}}+{\cal O}(\phi ^2)\,,
 \label{Wlinbeforegf}
\end{equation}
we get
\begin{equation}
U_{\underline{B}}{}^{\underline{A}}=\delta _{\underline{B}}{}^{\underline{A}}+ \phi_{\underline{B}}{}^{\underline{A}}+ {\cal O}(\phi ^2) \,,
 \label{Ulinbeforgf}
\end{equation}
After gauge fixing this is
\begin{equation}
  W=\unity _{32} +\frac{1}{2} \phi _a{}^{\dot a}\Gammabig ^a \Gammabig_{\dot a} + {\cal O}(\phi ^2) \,,
 \label{Wlinear}
\end{equation}
and we get
\begin{align}
U_{\underline{B}}{}^{\underline{A}}=&\delta _{\underline{B}}{}^{\underline{A}}+ \phi _a{}^{\dot a}\left(\delta _{\underline{B}}{}^a \delta ^{\underline{A}}_{\dot a}- (\eta _d) ^{a\underline{A}}(\eta _d)_{\dot a\underline{B}}\right)+ {\cal O}(\phi ^2)\nonumber\\
=&\delta _{\underline{B}}{}^{\underline{A}}+ \phi _a{}^{\dot a}\left(\delta_{\underline{B}} ^a\delta ^{\underline{A}}_{\dot a}+ \delta  ^{ a\underline{A}}\delta _{\dot a\underline{B}}\right)+ {\cal O}(\phi ^2) \,,
 \label{Ulinear}
\end{align}
which agrees with the linear part in (\ref{gf}), and is the linear part of the exponential of the fields times the $\SO(5,5)$ representation matrices.

This then implies for the $16\times 16$ matrices
\begin{align}
  V=\unity _{16} + \frac{1}{2}\phi^{ a\dot a}\Gamma '{}_{a \dot a}+ {\cal O}(\phi ^2) \,,\qquad
  \bar V =  &\unity _{16} + \frac{1}{2}\phi^{ a\dot a}\Gamma _{a \dot a}+ {\cal O}(\phi ^2)\,,
  \label{VbarVlinear}
\end{align}
where the $\Gamma$ are defined in (\ref{represSO55}) .

\subsection{Iwasawa  D+1 gauge}\label{Sec:parab}
 In terms of Fig. \ref{dynkin} this is the parabolic group $P_{\alpha_5}$ when the $\alpha_5$ node is removed. This gauge is related to the field content originating  from 7D  supergravity.

We are looking at the ungauged supergravity  in \cite{Tanii:1984zk,Bergshoeff:2007ef} trying to find a gauge in which the scalar part  becomes the one derived from 7D in  \cite{Cowdall:1998rs}. It was derived  from 7D gauged maximal supergravity in \cite{Pernici:1984xx} by a reduction to 6D and taking the limit to $g\to 0$.
The  6D ungauged supergravity action in
 \cite{Cowdall:1998rs} still has a local $SO(5)$ symmetry and  35=10+1+24 scalars are :
\be
 B^{I J}, \, {\sigma},\, {\Pi}_{I}^{\; \; j}
\label{Colscalars}\ee
Here $I=1,\dots, 5$, $i=1,\dots, 5$.
We will present the scalar part of the action in eqs. \rf{Cow},  \rf{Cow1}.
The action in  \cite{Cowdall:1998rs} has $SL(5, \mathbb{R})$  symmetry with 24 generators, inherited from 7D. The 10 axions $B^{I J}$ originate from 7D vector $A_7^{IJ}=B^{I J}$, $\sigma$ is related to an extra circle, and 24 scalars ${\Pi}_{I}^{\; \; j}$ represent a 7D
$SL(5, \mathbb{R})/SO(5)$ vielbein.

The local $SO(5)$ symmetry can be gauge-fixed,  so that only 25 physical scalars remain, these are $ B^{I J}, \, {\sigma}$ as well as 14 representatives of the coset space ${SL(5, \mathbb{R})\over SO(5)}$.
An Iwasawa triangular gauge in 6D supergravity I can be given in the form \rf{solpar}
where  25 generators of $SO(5,5)$ form a solvable Lie algebra.
 The representative of the coset $SO(5,5)/(SO(5)\times SO(5)$ is then
\be
  V_{\mu\dot \mu}{}^{\alpha\dot \alpha} (x)\qquad \Rightarrow \qquad V_{g.f}(x) = \exp \{ {B^{IJ} N_{IJ}  } \} {\Pi}^{Iwa} \exp \{\sigma h_0\}
\label{V16inNH}
 \ee
 where
 \be
{\Pi}^{Iwa} = \exp \{ \phi^{ij}t_{ij} +\phi^\lambda h_\lambda \}
\label{Iwa} \ee
 Here we have a set of 10 nilpotent generators, $N_{IJ}$, associated with axions  $B^{I J}$
  and a  set
 of 14 generators in \rf{Iwa}, representative of a  coset space ${SL(5, \mathbb{R})\over SO(5,\mathbb{R})}$ gauge-fixed in the Iwasawa gauge in ${\Pi}^{Iwa}$. These include
  10 nilpotent generators in $SL(5, \mathbb{R})$ $t_{ij}, i,j=1,..5, i<j$, and  4 non-compact Cartan generators of $SL(5, \mathbb{R})$,   $h_\lambda, \lambda=1, .. ,4$,  so this is the Borel subalgebra.
 There is also $h_0$,  the generator in $SO(5,5)$ which commutes with $SL(5,\mathbb{R})$.
Using this gauge allows to make a   7D origin of the 6D  maximal supergravity transparent.
\be
  V_{\mu\dot \mu}{}^{\alpha\dot \alpha} (x)\qquad \Rightarrow \qquad V_{g.f}(x) = \exp \{ {B^{IJ} N_{IJ}  } \} \exp \{ \phi^{ij}t_{ij} +\phi^\lambda h_\lambda \} \exp \{\sigma h_0\}
\label{para}
 \ee

\subsection{Partial Iwasawa D+1 gauge} \label{Sec:partial}

\noindent This gauge also has a 7D origin. To gauge-fix the local $SO(5)$ symmetry of the 6D maximal supergravity related to \cite{Cowdall:1998rs}  we proceed following the strategy in 4D maximal supergravity which we already applied in the partial Iwasawa
 gauge above.
We have to replace the  $SL(5, \mathbb{R})$ matrix ${\Pi}_{I}^{\; \; j}$ with 24-entries by a coset space ${SL(5, \mathbb{R})\over SO(5,\mathbb{R})}$ representative in a symmetric gauge,
so that with account of gauge-fixed local $SO(5)$ only 14 entries are left.  These are given in \cite{Gilmore:2008zz}.

The coset we need is described by the exponential of a hermitian symmetric (i. e. real) matrices $e^{H}$ such that
\be
H^\dagger = H\, ,\quad H^T= H\, , \quad \Tr H=0
\label{Hs}\ee
and
\be
(e^H)^\dagger = e^{H^\dagger}= e^H
\qquad
(e^H)^T = e^{H^T}= e^H
\label{H1s}\ee
These matrices are unimodular
\be
|| e^H|| = e^{\Tr H} =1
\label{H2s}\ee
Thus, the coset representative of ${SL(5, \mathbb{R})\over SO(5,\mathbb{R})}$ consist of all real symmetric unimodular matrices.
Once we constrain $\Pi$ as in eqs.  \rf{H1s}, \rf{H2s}
\be
\Pi  \qquad \to  \qquad \Pi = \Pi ^{\dagger} =\Pi^T\, , \quad || \Pi || =1
\label{Pi}\ee
we get a gauge-fixed unitary action with 25 scalars (14+10+1) and the remaining local $SO(5)$ symmetry inherited from 7D is fixed.

We can impose
 a symmetric gauge   associated with 14 scalars in ${SL(5, \mathbb{R})\over SO(5,\mathbb{R})}$
 \be
{\Pi}^{sym} = \exp \{ \phi^{ab} K_{ab} \}
\label{part} \ee
 where $K_{ab}$ are the coset generators and  $\Tr \phi=0$.  Note that in the Iwasawa gauge \rf{V16inNH} we had instead  ${\Pi}^{Iwa} $ in
\rf{Iwa}.
 Thus the total vielbein in partial Iwasawa D+1 gauge  is
\be
  V_{\mu\dot \mu}{}^{\alpha\dot \alpha} (x)\qquad \Rightarrow \qquad V_{g.f}(x) = \exp \{ {B^{IJ} N_{IJ} } \}
 \exp \{ \phi^{ab} K_{ab} \}
  \exp \{ {\sigma h_0 }\}
  \label{V16inNKh}
 \ee
and we get a gauge-fixed unitary gauge  with 25 scalars (14+10+1) and the remaining local $SO(5)$ symmetry fixed. Using this gauge allows to make a   7D origin of the 6D  maximal supergravity transparent. This is to be compared with Iwasawa D+1 gauge in  \rf{para}.

$SL(5, \mathbb{R})$ acts by left multiplication on the vielbein (\ref{V16inNKh}) together with a compensating
$SO(5)$ rotation from the right. So it rotates the $B^{IJ}$ (as it does in 7D), and acts non-linearly on the $\phi^{ab}$
via the standard coset action.

$E_{5(5)}$ is not manifest, but still present on the equations of motion (not the action),
it acts by left multiplication on the vielbein (\ref{V16inNKh}) together with a compensating
$SO(5)\times SO(5)$ rotation from the right. Such that this action is non-linear and in general mixes the $B^{IJ}$ with the $\phi^{ab}$.

\section{Supergravity I and   supergravity II in 6D}\label{Sec:7D6D}
The standard  6D ungauged maximal (2,2) supergravity with ${\cG\over \cH}={E_{5(5)}\over USp(4)\times USp(4)} \sim {SO(5, 5)\over SO(5) \times SO(5)}$ coset space was constructed in \cite{Tanii:1984zk,Bergshoeff:2007ef}.
It is presented  in the form with local  $\cH=SO(5) \times SO(5) $
symmetry. Duality symmetry of Gaillard-Zumino type  is realized acting on a 3-forms and
 their magnetic duals,    they combine into the vector representation 10 of SO(5,5)$\sim$ \EE\,.    A detailed version of this theory, in slightly different notation, as well as a gauged version of it was presented in  \cite{Bergshoeff:2007ef}. In this version  a $GL(5)$ electric subgroup of \EE\, is realized off-shell as a symmetry of the action.

Another version of 6D maximal supergravity  was derived in \cite{Cowdall:1998rs} from 7D gauged maximal supergravity in \cite{Pernici:1984xx}.  First, in
 \cite{Cowdall:1998rs} the gauged theory was reduced to 6D. The gauged version has $SO(5)_c\times SO(5)_g$ local symmetry. Here the local $SO(5)_c$ originates from the 11D Lorentz symmetry whereas $SO(5)_g$ is a Yang-Mills local symmetry due to gauging.
In the limit to ungauged 6D supergravity  in \cite{Cowdall:1998rs} the only local symmetry left is $SO(5)_c$ since the Yang-Mills $SO(5)_g$ disappears in the ungauged theory.

 The model with maximal local supersymmetry in 6D presented in eqs. (C.1)-(C.14) in \cite{Cowdall:1998rs} has the following properties originating from 7D theory in \cite{Pernici:1984xx}, which has  scalars associated  with the coset space ${SL(5, \mathbb{R})\over SO(5)}$.  In 7D the limit $g\to 0$ removing the $SO(5)_g$ gauge symmetry is singular.
Meanwhile in 6D the limit of removing the $SO(5)_g$ gauge symmetry is regular \cite{Cowdall:1998rs}. It has manifest linearly realized global $SL(5, \mathbb{R})$ symmetry, it has an independent local $SO(5)_c$ symmetry, both originating from D7 theory in \cite{Pernici:1984xx} after reduction on a circle and after  gauge coupling $g$ is sent to zero.

The field
content of the theory in eq. (C.2) in  \cite{Cowdall:1998rs} is: 1 graviton $g_{\mu\nu}$, 5 two-index antisymmetric tensor potentials
$C_{\mu\nu I}$, (10+5+1) vectors ($B_{\mu I}^{\; \; \; \; J}, S_{\mu
I}, A_{\mu}$),
4 gravitini ${\psi}_{\mu}$ and (16+4) spin 1/2 fermions
(${\lambda}_{i}, {\chi}$). All fermions ${\psi}_{\mu}$, ${\lambda}_{i}$ and ${\chi}$
are all D=6 USp(4) symplectic Majorana spinors.
The scalars are
\be
{\Pi}_{I}^{\; \; j}, \, B_{0I}^{\; \; \; \; J},\,
{\sigma}
\ee  originating from a reduction of D7 theory  \cite{Pernici:1984xx} on a $S^1$ circle. The 10 axions $B^{I J}$ originate from 7D vector $A_7^{IJ}=B^{I J}$, $\sigma$ is related to an extra circle, and 24 scalars ${\Pi}_{I}^{\; \; j}$ represent a 7D
$SL(5, \mathbb{R})$ vielbein.

${\Pi}_{I}^{\; \; j}$ is an $SL(5, \mathbb{R})$ matrix and has 24 entries. Here $i, j=1,\dots ,5$ are vector indices of $SO(5)_c$, which is still a local symmetry with 10 local parameters in the Lagrangian (C.2) of \cite{Cowdall:1998rs}. The indices $I, J= 1,\dots ,5$ originate from  a $SO(5)_g$ which is left after the limit of the Yang-Mills gauge coupling was taken. Thus the total amount of scalars, physical and unphysical in 6D maximal supergravity   in eq. (C.2) of \cite{Cowdall:1998rs} is 35=24+10+1 and there are still 10 local parameters of $SO(5)_c$ symmetry which can be used to remove the unphysical scalars to get the 25 physical
(14+10+1) scalars.
A direct inspection of the 6D maximal supergravity  in eq. (C.2) in \cite{Cowdall:1998rs} shows the following facts.

 The scalar part of the action is
\be
{1\over e} \cL^{sc}_{_{7D6D}}= -{1\over2}e^{4\sigma\over\sqrt{10}}({\Pi}_{I}^{\; \; i}{\Pi}_{J}^{\;
  \; j} \partial_{\mu}B^{IJ})^2
  -{1\over2}({\partial}_{\mu}{\sigma})^{2}-P_{\mu ij}P^{\mu ij}
\label{Cow}\ee
where
\bea
   P_{\mu ij}={\Pi}^{-1 \; I}_{\; \; \; \; (i}{\partial}_{\mu}^{\ }{\Pi}_{Ij)}^{\ }
\label{Cow1}\eea
This action is obtained from the scalar action of \cite{Tanii:1984zk,Bergshoeff:2007ef} by partially gauge fixing the $SO(5,5)$ vielbein as
\be
 V_{g.f}(x) = \exp \{ {B^{IJ} N_{IJ} } \} \,\Pi\,
  \exp \{ {\sigma h_0 }\}
  \;,
  \label{V16partial}
 \ee
which upon further gauge fixing of the remaining $SO(5)$ yields \rf{para} or \rf{V16inNKh}.

To gauge-fix the local $SO(5)_{c}$ symmetry of the 6D maximal supergravity in \cite{Cowdall:1998rs}  we proceed following the strategy in 4D maximal supergravity. As we already discussed in the context of the Iwasawa gauge for the action  in  \cite{Tanii:1984zk} in Sec. \ref{Sec:parab} and partial Iwasawa gauge in
Sec. \ref{Sec:partial}
we have to replace the  $SL(5, \mathbb{R})$ matrix ${\Pi}_{I}^{\; \; j}$ with 24-entries by a coset space ${SL(5, \mathbb{R})\over SO(5,\mathbb{R})}$ representative,
so that with account of gauge-fixed local $SO(5)_{c}$ only 14 entries are left.

One choice for  ${\Pi}_{I}^{\; \; j}$ can be
as in  \rf{Iwa}, the other
as in \rf{part}.
Once we constrain $\Pi$ as shown above
we get a gauge-fixed unitary action with 25 scalars (14+10+1) and the remaining local $SO(5)_{c}$ symmetry of the action in \cite{Cowdall:1998rs} is gauge-fixed. It is likely  that,  $SL(5, \mathbb{R})$ symmetry is still present but nonlinearly realized. But this requires an additional investigation.
$SL(5, \mathbb{R})$ is still present after gauge fixing.
With regard to $E_{5(5)}$ it is   not manifest and has no obvious reason to be present in 6D maximal supergravity II as a  hidden symmetry, either before or after local $SO(5)_c$ gauge-fixing. However,
$E_{5(5)}$ might be realized on the equations of motion, not on the action.

 There are 45 scalars in ungauged maximal 6D supergravity in \cite{Tanii:1984zk,Bergshoeff:2007ef}  in the vielbein in 16x16 representation  $V_{16}$  or in 10x10 representation in $\cV_{10}$ of $SO(5, 5)$. In notation of \cite{Bergshoeff:2007ef} the 16x16 vielbein is $ V_{M}{}^{\alpha\dot \alpha} (x)$,  and an $SO(5) \times SO(5)$ covariant 1-form is
 \be
 P^{a\dot a} = {1\over 4} \bar V \gamma^a \gamma^{\dot a} d V
\ee
 and the scalar Lagrangian has a local       $SO(5) \times SO(5)$ symmetry and a global \EE\, symmetry
\be
{1\over e} \cL^{sc}_{_{ 6D}}= -{1\over 16} P_\mu{}^{a\dot a} P^\mu{}_{a\dot a}
\ee
When the local $SO(5) \times SO(5)$ symmetry is gauge-fixed, the 20 local parameters can be used to eliminate 20 scalars so that only 25 physical scalars remain.

There are
 5 two-index antisymmetric tensor potentials $C_{\mu\nu\, I }$ and 16=10+5+1 vector fields $B_{\mu I}{}^J, S_{\mu I}, A_{\mu}$ in 6D ungauged supergravity in \cite{Cowdall:1998rs} with kinetic terms depending on 35 scalars, or on 25 after the local $SO(5)_{c}$ gauge symmetry is gauge-fixed as in \rf{Iwa} or \rf{part}.

The maximal 6D supergravity action in eq. (C.2) in \cite{Cowdall:1998rs} has the following  action for the tensors and vectors
\bea
  e^{-1}{\cal L}_{7D6D}^{ten, vec} &=& -{1\over4}e^{-{5\sigma\over\sqrt{10}}}(f_{\mu\nu})^2
  -{1\over{12}}e^{-{2\sigma\over\sqrt{10}}}({\Pi}^{-1 \; \; I}_{\; \; \;
  \; i}H_{\mu\nu\rho I})^2
  -{1\over4}e^{-{\sigma\over\sqrt{10}}}({\Pi}_{I}^{\; \; i}{\Pi}_{J}^{\;
  \; j} F_{\mu\nu}^{IJ})^2
\nonumber\\
&& -{1\over{4}}e^{3\sigma\over\sqrt{10}}({\Pi}^{-1 \; \; I}_{\; \; \;
  \; i}G_{\mu\nu I})^2
  \nonumber\\
&&-{{e^{-1}}\over36\sqrt{2}}{\epsilon}^{\mu\nu\rho\sigma\lambda\tau}
B_{0}^{\; \; IJ}H_{\mu\nu\rho I}H_{\sigma\lambda\tau J}
-{{e^{-1}}\over6\sqrt{2}}{\epsilon}^{\mu\nu\rho\sigma\lambda\tau}
H_{\mu\nu\rho I}B_{\sigma}^{\; \; IJ}G_{\lambda\tau J}
\label{Cowdall}\eea
 Here
 \bea
&f_{\mu\nu} \rightarrow   f_{2}=dA \; \; \; \; \; \; \; \; G_{\mu\nu I} \rightarrow G_{2I}=dS_{1I}
\nonumber\\
&F_{\mu\nu}^{IJ}\rightarrow   F_{2 I}^{\; \; \; J}= dB_{1 I}^{\; \; \; J}+B_{0 I}^{\; \; \;
   J}dA \; \; \; \; \; \; \; \nonumber\\
&  {H}_{\mu\nu\rho I} = 3({\partial}_{[ \mu}C_{\nu\rho ]I}
  +{1\over2}{G}_{[ \mu\nu I}{A}_{\rho ]})
\eea
By comparison, the tensor-vector part of the 6D action in \cite{Tanii:1984zk,Bergshoeff:2007ef} is
\be
{1\over e} \cL^{ten, vec}_{_{ 6D}}=   -\ft1{12} H_m \cdot K^{mn} H_n -\ft14 \cM_{AB} F_{\mu\nu}^\cA F^{\mu\nu\cB}
+ H_m \cdot j\omega^m -\ft1{12}\omega_m\cdot j \omega^m \ ,\label{LTanii}
\ee
The action in  \cite{Tanii:1984zk} is the same as the $g\to 0$ in   \cite{Bergshoeff:2007ef}, in slightly different notations. It has
$SO(5)\times SO(5)$ local symmetry and an on shell \EE\,  global symmetry.  We refer to notations and details in   \cite{Bergshoeff:2007ef}.

Thus the unitary gauge of the action in \rf{Cowdall} is established. It is hard to see where is the global $SO(5)\times SO(5)$ and reflection symmetries in this case. Therefore the absence of 1-loop anomalies for these symmetries found in  \cite{Marcus:1985yy} does  not make sense here, unless one can prove that this unitary gauge is equivalent to a symmetric gauge in  \cite{Tanii:1984zk}.  It is also not clear how this 6D supergravity theory II can be protected from UV divergences and  anomalies.

\section{Discussion}

The purpose of this work was to study  gauge-fixing of local H symmetries in G/H maximal D-dimensional supergravities. The existence of different unitary gauges in G/H maximal D-dimensional supergravities
has a deep origin in existence of different versions of D-dimensional supergravities, some of which are associated with D+n-dimensional supergravities. We studied mostly the case n=1, i. e. D+1 supergravities compactified on a circle. In certain gauges, like symmetric gauges,  all scalars enter the action non-polynomially. In other gauges, like Iwasawa gauges,   associated with D+1-dimensional supergravities, which we studied here, some of the scalars enter the action polynomially.

The reason for this difference has to do with the properties of the Lie algebra of the relevant duality $G_U=E_{11-D(11-D)}$ symmetries of D-dimensional maximal supergravities \cite{Andrianopoli:1996zg}. There is a
{\it Cartan decomposition} of a duality symmetry  Lie algebra
\be
\mathbb{G}= \mathbb{H}\oplus \mathbb{K}
\ee
Here $\mathbb{H}$ is a compact subalgebra of $\mathbb{G}$ and $\mathbb{K}$ includes generators of the  coset space G/H. For example, in $\cN=8$ in 4D  there are 63 generators in   $H=SU(8)$ and 70  in ${E_{7(7)}\over SU(8)}$. In symmetric gauges global H-symmetry is preserved and supergravity vielbeins take the form
\be
{\cal V}_{sym}= e^{\phi_{sym} \cdot \mathbb{K}}
\ee
For symmetric gauge to be preserved under G-symmetry transformation a compensating field-dependent H-symmetry transformation is required. As the result, the global G-symmetry remains a symmetry of the theory and it is non-linearly realized. Therefore the studies of the vanishing soft scalar limits are efficient and useful in amplitude computations.

\noindent There is an  {\it Iwasawa decomposition} of a duality symmetry  Lie algebra
\be
\mathbb{G}= \mathbb{H}\oplus \mathscr {S}\ee
Here $\mathscr {S}$ is a solvable subgroup of the Lie algebra,
\be
 \mathscr {S}={ C}\oplus {  N}\,,\label{SCN2}
\ee
${ C}$ is the Cartan subspace of the coset space and  $N$ is a nilpotent subalgebra. Since ${ N}$ is nilpotent, {\it the axionic scalars parametrizing ${ N}$ appear in the supergravity action
polynomially}. In Iwasawa-type  gauges global H-symmetry as well as G-symmetry are either broken or, at least,  not manifest
\be
{\cal V}_{Iwa}= e^{\phi_{Iwa} \cdot  \mathscr {S}}
\ee
In Iwasawa gauges the scalars are the  same as in  supergravity theories II which were derived by a compactification from D+1 dimension where scalars are in a coset space $(G/H)_{D+1}$.  In these versions of D-dimensional supergravity there is a smaller local symmetry $H_{D+1} < H_D$ as well as a smaller global symmetry $G_{D+1} < G_D$.

Thus, the existence of different D-dimensional supergravities, I and II,  and different gauges for a local H-symmetry in supergravities I is universal, and we studied all of this in various examples, particularly in 4D and 6D.
The next question raised by our investigation here is the relation between different local H-symmetry gauges in supergravities I, i. e. the issue of local H-symmetry anomalies.  Another question is the relation between different versions of supergravities, I and II,  which involves the issue of G-symmetry anomalies since in supergravities II
$G_D$-symmetry is broken down to $G_{D+1}$ by the Group Disintegration process \cite{Hawking:1981bu}.

 In 4D it was established in a set of papers in  \cite{deWit:2002vt,deWit:2005ub,deWit:2007kvg} that using an $Sp(56, \mathbb{R})$ Gaillard-Zumino duality symmetry one can show, for example, that starting with supergravity I in \cite{Cremmer:1979up,deWit:1982bul} one can  change a symplectic frame from the $SL(8,\mathbb{R})$-basis to the   $E_{6(6)}$-basis and  gauge-fix local symmetries  in a way which allows to reproduce the results of supergravity II in  \cite{Andrianopoli:2002mf} for physical observables. This suggests that the extra symmetries beyond U-duality may help to understand the relation between different version of D-dimensional supergravities and different gauges, which in turn might be helpful for understanding UV divergences in D-dimensional supergravities \cite{Kallosh:2024ull}.

 \

\noindent{\bf {Acknowledgments:}}  We had helpful suggestions  from P. Aspinwall, E. Bergshoeff,  G. Moore and M. G\"unaydin  concerning the issue  of the 16x16 representative of the coset space  of $SO(5,5) \over SO(5)\times SO(5)$ which is not  standard, as opposite to the 10x10 representation available in the textbooks.
We are grateful to  participants of the Amplitudes 2024 conference for the interest to this work, especially to N. Arkani Hamed, L. Dixon, A. Edison, H. Johansson, J. Parra-Martinez and R. Roiban. The discussion with them suggested that the current work poses the following questions. Which of these various D-dimensional supergravities, quantized in different gauges, are described by the loop computations using the  amplitude double-copy methods? How to probe  $Sp(2n)$ symmetry in amplitudes? We also thank C. Hull, M. Henneaux and the organizers of the Mons workshop `Gauge invariance: quantization and geometry- in memory of Igor Batalin' for discussions.

 The work of RK is supported by SITP and by the US National Science Foundation grant PHY-2310429.

\appendix
\section{6D Notation}
\label{app:gammas}
Our notations are consistent with \cite{Freedman:2012zz}, and mostly with  \cite{Tanii:1984zk,Bergshoeff:2007ef,Karndumri:2021llj}.
We start with a table of the indices:
\begin{align}
 & a,\dot a  = 1,\ldots 5\,,\qquad \underline{A}=\{a,\dot a\}=1,\ldots ,10\,, \nonumber\\
&  m=1,\ldots 5\,,\qquad M=1,\ldots ,10\,, \nonumber\\
&  \alpha ,\dot \alpha =1,\ldots ,4\,,\qquad A=(\alpha \dot\alpha )=1,\ldots ,16  \,,\nonumber\\
&  i=1,2,\qquad {\cal A}=(\dot \alpha \alpha  i)= (Ai)= 1,\ldots ,32\,.
  \label{spinorindices}
\end{align}
The full $\SO(5,5)$ gamma matrices are written as
\begin{equation}
  (\Gammabig _a)_{\cal A}{}^{\cal B} = (\gamma _a)_a {}^\beta \delta _{\dot a }{}^{\dot \beta }(\sigma _1)_i{}^j\,,\qquad (\Gammabig _{\dot a})_{\cal A}{}^{\cal B} = \delta_a {}^\beta (\gamma _{\dot a})_{\dot \alpha }{}^{\dot \beta }(\rmi\sigma _2)_i{}^j\,,\qquad {\cal A}=(\dot \alpha \alpha  i)\,,\quad {\cal B}= (\dot \beta \beta  j)\,
 \label{Gammaadota}
\end{equation}
Here one starts from the $D=5$ Euclidean gamma matrices $\gamma _a$ and $\gamma _{\dot a}$, which are identical:
\begin{align}
  &\gamma _a\gamma _b+\gamma _b\gamma _a=2\delta _{ab}\unity _4 \,,\qquad \gamma _{\dot a}\gamma _{\dot b}+\gamma _{\dot b}\gamma _{\dot a}=2\delta _{\dot a\dot b}\unity _4\,,\nonumber\\
  &\Gammabig _{\underline{A}}=\{\Gammabig _a,\,\Gammabig _{\dot a}\}\,,\qquad \Gammabig _{\underline{A}}\Gammabig _{\underline{B}}+\Gammabig _{\underline{B}} \Gammabig _{\underline{A}}= 2(\eta _d)_{\underline{A}\underline{B}}\,,\nonumber\\
  &\Gammabig _a\Gammabig _b+\Gammabig _b\Gammabig _a=2\delta _{ab}\unity _{32} \,,\qquad \Gammabig _{\dot a}\Gammabig _{\dot b}+\Gammabig _{\dot b}\Gammabig _{\dot a}=-2\delta _{\dot a\dot b}\unity _{32}\,,\qquad  \Gammabig _{a}\Gammabig _{\dot b}+\Gammabig _{\dot b}\Gammabig _{ a}=0\,.
  \label{gammaalgebraab}
\end{align}
Similarly, the vector indices are raised as
\begin{equation}
  \gamma ^a=\delta ^{ab}\gamma _b\,,\qquad \gamma ^{\dot a}= \delta ^{\dot a\dot b}\gamma _{\dot b}\,,\qquad
  \Gammabig ^{\underline{A}}= (\eta _d)^{\underline{A}\underline{B}}\Gammabig _{\underline{B}}=\{\Gammabig ^a,\,\Gammabig ^{\dot a}\}=\{\delta ^{ab}\Gammabig _b,\,-\delta ^{\dot a\dot b}\Gammabig_{\dot b}\} \,.
 \label{gammaup}
\end{equation}

Lowering spinor indices is done in 5D with $\Omega _{\alpha \beta }$ and in 10D with $\mathbb{C}_{{\cal A}{\cal B}}$
\begin{align}
  &(\gamma _a)_{\alpha \beta }= (\gamma _a )_\alpha {}^\gamma \Omega _{\gamma \beta }= -(\gamma _a)_{ \beta\alpha }\,,\qquad \Omega_{\beta \alpha } = -\Omega_{\alpha \beta }\,,\nonumber\\
 &( \Gammabig _{\underline{A}})_{{\cal A}{\cal B}}=( \Gammabig _{\underline{A}})_{{\cal A}}{}^{\cal C}\mathbb{C}_{{\cal C}{\cal B}}= ( \Gammabig _{\underline{A}})_{{\cal B}{\cal A}
 }\,,\qquad \mathbb{C}_{{\cal A}{\cal B}}= \Omega_{\dot \alpha \dot \beta } \times \Omega_{\alpha \beta } \times (\rmi\sigma _2)_{ij}= -\mathbb{C}_{{\cal B}{\cal A}}\,.
  \label{D5chargeconj}
\end{align}
Lowering only the $A$ index goes thus with
\begin{equation}
  C_{AB}=C_{(\alpha \dot \alpha)(\beta \dot \beta ) }=\Omega_{\dot \alpha \dot \beta } \times \Omega_{\alpha \beta }=C_{BA}\,,\qquad C^{AB}C_{CB}=\delta ^A_C\,.
 \label{CAB}
\end{equation}
and thus the value of $C_{AB}$ is the same as the value of $C^{AB}$. The relation can also be written as
\begin{align}
  &\mathbb{C}_{{\cal A}{\cal B}}=C_{AB}\varepsilon _{ij}=\begin{pmatrix}0&C_{AB}\cr -C_{AB}&0\end{pmatrix}_{ij} \qquad  \mbox{for }{\cal A}=(Ai),\,{\cal B}=(Bj)  \,,\nonumber\\
  &\mathbb{C}^{{\cal A}{\cal B}}=C^{AB}\varepsilon ^{ij}=\begin{pmatrix}0&C^{AB}\cr -C^{AB}&0\end{pmatrix}_{ij}\,,\qquad \mathbb{C}^{{\cal A}{\cal C}}\mathbb{C}_{{\cal B}{\cal C}}=\delta^{{\cal A}}_{{\cal C}}\,.
  \label{chargeconj3216}
\end{align}

We can define
\begin{align}
  &\Gammabig _*= \Gammabig _{\dot 1}\Gammabig _{\dot 2}\Gammabig _{\dot 3}\Gammabig _{\dot 4}\Gammabig _{\dot 5}\Gammabig _1\Gammabig _2\Gammabig _3\Gammabig _4\Gammabig _5 = \unity _{16}\times \sigma _3\,,\nonumber\\
 &P_L = \frac{1}{2}(\unity _{32}+\Gammabig _*)=\begin{pmatrix}\unity _{16}&0\cr 0&0\end{pmatrix}\,,\qquad P_R = \frac{1}{2}(\unity _{32}-\Gammabig _*)=\begin{pmatrix}0&0\cr 0&\unity _{16}\end{pmatrix}\,.
\label{Gamma*}
\end{align}

Making the $i$ index explicit, we can also define $16\times 16$ matrices $\Gamma _{\underline{A}}$ and $\Gamma '_{\underline{A}}$ as chirally projected matrices from $\Gammabig _{\underline{A}}$:
\begin{align}
&\Gammabig _{\underline{A}}=\begin{pmatrix}0&\Gamma_{\underline{A}}\cr \Gamma '_{\underline{A}}&0\end{pmatrix}\,,\qquad \Gamma _{\underline{A}}=\{\Gamma _a,\Gamma _{\dot a}\}\,,\qquad\Gamma' _{\underline{A}}=\{\Gamma' _a,\Gamma' _{\dot a}\}\,,\nonumber\\
&\Gammabig ^{\underline{A}}=\begin{pmatrix}0&\Gamma^{\underline{A}}\cr \Gamma '{}^{\underline{A}}&0\end{pmatrix}\,,\qquad \Gamma ^{\underline{A}}=(\eta _d)^{\underline{A}\underline{B}}\Gamma _{\underline{B}}=\{\Gamma ^a,\Gamma ^{\dot a}\}\,,\qquad\Gamma'{} ^{\underline{A}}=(\eta _d)^{\underline{A}\underline{B}}\Gamma'{}_{\underline{B}}=\{\Gamma'{}^a,\Gamma'{} ^{\dot a}\}\,,\nonumber\\
&\Gamma _a=\Gamma '_a= \unity _4\times \gamma_a\,,\qquad
\Gamma _{\dot a}=-\Gamma '_{\dot a}=\gamma_{\dot a}\times\unity _4\,,\nonumber\\ &\Gamma ^a=\Gamma '{}^a= \unity _4\times \gamma^a\,,\qquad
\Gamma ^{\dot a}=-\Gamma '{}^{\dot a}=-\gamma^{\dot a}\times\unity _4\,,
\,.
 \label{PLRGamma}
\end{align}
Note that $\Gamma'{} ^{\underline{A}}=\Gamma _{\underline{A}}$.
Their basic relations follow from  (\ref{gammaalgebraab}):
\begin{align}
  \Gamma_{\underline{A}}\Gamma'_{\underline{B}}+  \Gamma_{\underline{B}}\Gamma'_{\underline{A}}= 2(\eta _d)_{\underline{A}\underline{B}}\,.
  \label{gammaunderlinealg}
\end{align}
Furthermore, the trace rules for the $16\times 16$ matrices is
\begin{equation}
  \Tr(\Gamma _{\underline{A}}\Gamma '{}^{\underline{B}})= \Tr(\Gamma' _{\underline{A}}\Gamma ^{\underline{B}}))= 16\delta _{\underline{A}}^{\underline{B}}\,.
 \label{Trace16}
\end{equation}

The $\Gamma _{\underline{A}}$ (and their primed versions) have naturally indices $A,B$: $(\Gamma _{\underline{A}})_A{}^B$, and if the latter is lowered they are symmetric:
\begin{equation}
 (\Gammabig _{\underline{A}})_{(Ai)(Bj)}=\begin{pmatrix}-(\Gamma _{\underline{A}})_{AB}&0\cr 0&(\Gamma' _{\underline{A}})_{AB}\end{pmatrix}_{ij}\,,\qquad
  (\Gamma _{\underline{A}})_{AB}=  (\Gamma _{\underline{A}})_A{}^C C_{CB}= (\Gamma _{\underline{A}})_{BA}\,.
 \label{gammaunderlinesymmetric}
\end{equation}

The representation matrices of $\SO(5,5)$ are then
\begin{align}
 &\Gammabig _{\underline{A}\underline{B}}=\Gammabig _{[\underline{A}}\Gammabig_{\underline{B}]}=\begin{pmatrix} \Gamma _{\underline{A}\underline{B}} & 0 \cr 0& \Gamma' _{\underline{A}\underline{B}}\end{pmatrix}\,,\nonumber\\
&  \Gamma _{\underline{A}\underline{B}} = \frac{1}{2}(\Gamma _{\underline{A}}\Gamma '_{\underline{B}}- \Gamma _{\underline{B}}\Gamma '_{\underline{A}})\,,\qquad
  \Gamma'_{\underline{A}\underline{B}} = \frac{1}{2}(\Gamma' _{\underline{A}}\Gamma _{\underline{B}}- \Gamma '_{\underline{B}}\Gamma _{\underline{A}})\,, \nonumber\\
  &\Gamma _{ab}=\Gamma '_{ab}= \unity _4\times \gamma_{ab}\,,\qquad \Gamma _{\dot a\dot b}=\Gamma' _{\dot a\dot b}= -\gamma_{\dot a\dot b}\times\unity _4\,,\nonumber\\
&  \Gamma     _{a\dot a}=-\Gamma _{\dot aa}=-\gamma_{\dot a}\times\gamma _a\,,\qquad \Gamma '    _{a\dot a}=-\Gamma '    _{\dot a a}= \gamma_{\dot a}\times\gamma _a\,.
  \label{represSO55}
\end{align}

\section{Minimal unitary representations of $E_{7(7)}$ and $E_{5(5)}$  }\label{App:A}
In the context of  gauge-fixing local H-symmetry in supergravities  it is useful to consider also
{\it minimal unitary representations of the group} G  as presented in \cite{Joseph:1974hr}, \cite{Kazhdan:2001nx}
with regard to
quantum mechanical system which admit  a semisimple Lie algebra. Minimal realization is the case of   the least number of degrees of freedom for which a quantum mechanical system
admits  a given  Lie algebra. For semisimple algebra ${\mathfrak g}$ there is a  canonical decomposition
\be
{\mathfrak g}= {\mathfrak n}^+ \oplus {\mathfrak h}  \oplus {\mathfrak n}^-
\ee
where $ {\mathfrak h}$ is a Cartan subalgebra and ${\mathfrak n}^\pm $ are the nilpotent subalgebras spanned respectively by the positive and negative root eigenvectors.
The procedure of constructing  explicitly  the minimal unitary representations of a Lie group G was provided in details in \cite{Kazhdan:2001nx}, where the parabolic subgroups P of G
are involved.

The algebraic structure of quantum mechanics and the concept of a spectrum generating algebras emphasised in  \cite{Joseph:1974hr,Kazhdan:2001nx} are based on canonical commutation relations
 $[p_i, x^j]=-i \delta_i{}^j$ and $p_i$ is realized as $-i {\partial\over \partial x^i}$.
For example, the Dynkin diagram for $E_7$ in Fig. \ref{EA} corresponds to Cartan generators and simple roots presented in  \cite{Kazhdan:2001nx} in the form $H_{\beta_0}= -y\partial +x_0 \partial_0, \dots$ and $E_{\alpha_1}= x_2 \partial_3 +x_4 \partial_5 +x_6 \partial_8 +x_9 \partial_{12}, \dots$,  etc acting on $(y, x_0, x_1, \dots, x_{15})$.
An important role in this construction is played by a parabolic subgroups  of G where the right node in Dynkin diagrams in Figs. \ref{dynkin}, \ref{EA} is removed.

  The  common feature of  all unitary gauges in supergravities, symmetric, and  Iwasawa-type   is that the number of physical scalars is the same, e. g.  70 in 4D maximal supergravity and 25 in 6D maximal supergravity. However, the difference between these gauges is that only Iwasawa  gauges, where coset generators are in a subalgebra of G,  qualify for minimal unitary representations of the group G as defined in \cite{Joseph:1974hr,Kazhdan:2001nx}.

The explicit unitary representations for 6D maximal supergravity were given in \cite{Bossard:2014lra} and
for 4D maximal supergravity  in \cite{Bossard:2015uga}. They were obtained by solving differential equations defining representations of \E\, or \EE\,  in a parabolic gauge associated with the so-called decompactification limit.

Therefore these  minimal unitary representations for G-duality groups  of 4D and 6D maximal supergravities are represented by a field content of 5D and 7D supergravities, respectively, as we have explained in Secs. \ref{Sec:5D4D} and \ref{Sec:7D6D}. These minimal unitary representations  also represent  Iwasawa-type  class of gauges of a local H-symmetry.

{\it 4D maximal supergravity }

\noindent Minimal unitary representation is presented in the form of the  coset representative in a parabolic  gauge of 4D maximal supergravity in \cite{Bossard:2015uga}. It is based on a decomposition of the \, Lie algebra
\be
{\mathfrak e}_{7(7)}\cong \, \overline{ {\bf 27}}^{(-2)} \oplus  ({\mathfrak g} {\mathfrak l}_1\oplus {\mathfrak e}_{6(6)} )^0 \, \oplus \, {\bf 27}^{(2)}
\ee
This is the same as the one in 5D4D supergravity in \cite{Andrianopoli:2002mf}  where it was given in the form
\be
{\mathfrak e}_{7(7)}= {\mathfrak e}_{6(6)} + {\mathfrak s} {\mathfrak o}(1,1) +{\mathfrak p}\, , \qquad {\mathfrak p}= {\bf 27}_{(-2)} + {\bf 27}'_{(+2)}
\ee
so that  under the subgroup $E_{6(6)}\times SO(1,1)$ there is a decomposition
\be
{\bf 56} \rightarrow   {\bf
1}_{-3}+{\bf 27}'_{-1}+{\bf 1}_{+3} +    {\bf 27}_{+1} \,,
\label{56e6}
\ee
The 56x56-bein defining the minimal unitary representation of \E\, was given  in \cite{Bossard:2015uga} in the form
\be \cV_{56} = \left( \begin{array}{cccc} \ e^{3\phi} \ & \ 0 \ & \ 0 \ & \ 0 \\
0 &\ e^{\phi} V_{ij}{}^I \ &0&0\\
0&0& \ e^{-\phi} (V^{-1}){}_I{}^{ij} \ &0\\
0&0&0& e^{-3\phi} \end{array}\right) \left( \begin{array}{cccc} \ 1 \ & \ a^J \ & \ \tfrac{1}{2} t_{JKL} a^K a^L  \ & \ \tfrac{1}{3} t_{KLP} a^K a^L a^P \\
0 &\ \delta_I^J \ &t_{IJK} a^K&\tfrac{1}{2} t_{IKL} a^K a^L\\
0&0& \delta_J^I \ &a^I \\
0&0&0& 1 \end{array}\right)  \ , \ee
where $V_{ij}{}^I$ is a coset representative of $E_{6(6)} / USp(8)$, and $t_{IJK}$ is the invariant symmetric tensor of $E_{6(6)}$. Here $\cV_{56}$ depends on 70= 1+42+27 scalars.
\be
\phi\, , \, V_{ij}{}^I\, , \, a^I
\ee
This is in agreement with the construction in \cite{Andrianopoli:2002mf}.
It corresponds to a partial Iwasawa gauge we discussed above. One could have used the same expression for $\cV_{56}$ where $V_{ij}{}^I$ is a representative of $E_{6(6)} / USp(8)$ in a solvable parametrization, which would give an Iwasawa, triangular gauge for the 56-bein. This is what we did in Sec. \ref{Sec:4D}.

{\it 6D maximal supergravity }

\noindent
Here the minimal unitary representation of $E_{5(5)}$ in \cite{Bossard:2014lra} is based on a decomposition
\be
{\mathfrak e}_{5(5)}\cong \, \overline{ {\bf 10}}^{(-2)} \oplus  ({\mathfrak g} {\mathfrak l}_1\oplus {\mathfrak s}{\mathfrak l}
_{5} )^0 \, \oplus \, {\bf 10}^{(2)}
\ee
This  again can be related to 6D  supergravity II in \cite{Cowdall:1998rs} which we described in Sec. \ref{Sec:7D6D}.
\noindent The corresponding 10x10 and 16x16 matrices are given in \cite{Bossard:2014lra}
\be
{\cal V}_{10}=\left(\begin{array}{cc}e^{2\phi} (v^{-1})_J{}^a & \, \hskip 1 cm  e^{2\phi} (v^{-1})_K{}^a a^{KJ}\\
\\
0 & \, \hskip 0.6 cm e^{2\phi} v_a{}^J\end{array}\right)
\label{10}\ee
and
\be
{\cal V}_{16}=\left(\begin{array}{ccc}e^{5\phi} &\, \hskip 0.5 cm  {1\over \sqrt 2} e^{5\phi}  a^{KL} &\, \hskip 0.5 cm  {1\over 8} e^{5\phi} \epsilon_{KPQRS} a^{PQ} a^{RS}\\
\\
0 &\, \hskip 0.2 cm  e^\phi v_{[a}{}^K v_{a]}{}^L &\, \hskip 1 cm  e^\phi v_{[a}{}^R v_{a]}{}^S {1\over 2\sqrt 2}  \epsilon_{RSKPQ} a^{PQ} \\
\\
0 & 0 & e^{-3\phi} (v^{-1})_K{}^a\end{array}\right)
\label{16}\ee
Matrices ${\cal V}_{10}$ and ${\cal V}_{16}$ have 35 independent entries
\be
\phi\, , \quad v_a{}^J\, , \quad a^{KJ}
\label{35}\ee
Here the 24 scalars $v_a{}^J$ are in $SL(5, \mathbb{R})$ and there is still one local $SO(5)$ symmetry in the theory which has to be gauge-fixed. Therefore to bring these expressions, depending on $1+24+10=35$ scalars to the ones which depend only on 25 physical scalars we have to gauge-fix the  local $SO(5)$ by taking a coset representative for ${SL(5, \mathbb{R})\over SO(5)}$. This means replacing the 24 scalars in $v_a{}^J$ by the coset representative of ${SL(5, \mathbb{R})\over SO(5)}$ which depends on 14 scalars.  This can be done either in  a solvable parametrization or in a symmetric gauge. In this way one can fully specify the 10-bein and the 16-bein to depend on 25 scalars either in Iwasawa, triangular gauge or in a partial Iwasawa gauge as we did above in Sec. \ref{Sec:6D}.

\section{Action of maximal 6D supergravity and amplitudes}

The action in  \cite{Tanii:1984zk} is the same as the $g\to 0$ in   \cite{Bergshoeff:2007ef}, in slightly different notations. It has
$SO(5)\times SO(5)$ local symmetry and an on shell \EE\,  global symmetry. We have shown  bosonic actions in Sec. \ref{Sec:7D6D}. The  fermionic kinetic terms in Sec. (2.6) of   \cite{Bergshoeff:2007ef} are
\bea
{1\over e} {\cal L}^{ferm}_{_{6D}} =
- \ft12 {\bar\psi}_{+\mu} \c^{\mu\nu\rho} D_\nu \psi_{+\rho}
-\ft12 {\bar\psi}_{-\mu} \c^{\mu\nu\rho} D_\nu \psi_{-\rho}
-\ft12 {\bar\chi}^a \c^\mu D_\mu \chi^a
-\ft12 {\bar\chi}^\adot\c^\mu D_\mu \chi^\adot
\label{LTanii2}
\eea
and we skipped terms with fermion interaction with bosons.
The spinor fields are   $\psi_{+ \mu\alpha}, \psi_{- \mu\dot \alpha}$  and  $\chi_{+ a \dot \alpha}, \chi_{- \dot a  \alpha}$ where $a, \dot a =1,...,5$ and $\alpha, \dot \alpha$ label spinors of $SO(5)\times SO(5)$. Here $\pm$ refers to spacetime chirality of the spinors which are 6D symplectic Majorana-Weyl.

When this action is gauge-fixed in a symmetric gauge it is a model  for which anomaly computation in  \cite{Marcus:1985yy} is relevant.
Note that the action is manifestly invariant under reflection: flipping chirality and $SO(5)_1$ to $ SO(5)_2$.
This symmetry plays an important role in cancellation of 6D supergravity $SO(5)\times SO(5)\sim USp(4)\times USp(4)$ anomalies in  \cite{Marcus:1985yy}.

The maximal 7D6D ungauged supergravity action in eq. (C.2) in \cite{Cowdall:1998rs} has $SO(5)$ local symmetry. We have shown  bosonic terms in the action in Sec. \ref{Sec:7D6D}.
The  fermionic kinetic terms are
\bea
  {1\over e} {\cal L}^{ferm}_{_{7D6D}} =
-{\bar\psi}_{\mu}{\tau}^{\mu\nu\rho}{\nabla}_{\nu}{\psi}_{\rho}
   -{\bar\chi}{\tau}^{\mu}{\nabla}_{\mu}{\chi}
   -{\bar\lambda}^{i}{\tau}^{\mu}{\nabla}_{\mu}{\lambda}_{i} +\dots
\label{Cowdall2}\eea
where the $\dots$ involve terms with fermion interaction with bosons.  The fermions are $USp(4)$ 6D Majorana symplectic spinors, they are Lorentz  and $SO(5)$-covariant. This action has no manifest symmetry under refection when flipping chirality and $USp(4)_1$ to $ USp(4)_2$.

In symmetric gauge in  \cite{Tanii:1984zk,Bergshoeff:2007ef} there are
25 scalars $\phi^{a\dot a}$ with $a, \dot a=1,\dots, 5$  viewed  with spinorial indices in $USp(4)\times USp(4)$ are
\be
W_{\alpha \beta}^{\dot \alpha \dot \beta}(x) \equiv  (\gamma_a)_{\alpha \beta} (\gamma_{\dot a})^{\dot \a \dot \beta} \phi^{a\dot a}(x)\, , \qquad \alpha, \dot \alpha=1,2,3,4
\ee
At the linear level these scalars are the first components in  the linearized BPS superfields $W_{\alpha \beta}^{\dot \alpha \dot \beta}(x, \theta)$ \cite{Kallosh:2023dpr,Kallosh:2024lsl}.
For example  $ W_{12}^{\dot 1 \dot 2}(x, \theta) $ is a 1/2 BPS superfield, it  depends on half of fermionic directions in superspace
\be
D_{\a 1} W_{12}^{\dot 1 \dot 2} = D_{\a 2} W_{12}^{\dot 1 \dot 2} = D^{\dot \a \dot 1} W_{12}^{\dot 1 \dot 2} =D^{\dot \a \dot 2} W_{12}^{\dot 1 \dot 2}=0   \ .
\label{12BPS}\ee

Using superamplitudes, the structure of the maximal supergravity 4-point tree amplitude was given in  \cite{Cachazo:2018hqa}. The corresponding  on shell superfield   depends on 8 Grassmann coordinates
and it is directly related to $ W_{12}^{\dot 1 \dot 2}(x, \theta) $.
The four-point superamplitude is given in \cite{Cachazo:2018hqa} in the form
\be \label{eq:sugra-4pts}
{M}_4^{\N=(2,2)\text{ tree}} \;=\; {1\over \kappa^2} \d^6 \left( \sum_{i=1}^4 p_i^{AB} \right) { \delta^{8} \left( \sum_{i=1}^4 q^{A, I}_i \right)  \delta^{8} \left( \sum_{i=1}^4 {\tilde q}^{ {\hat I} }_{i,{\hat A}}\right)  \over s_{12}\, s_{23}\, s_{13} } \, ,
\ee
We have  presented in \cite{Kallosh:2023dpr}  a local linearized superinvariant defining  3-loop 4-point UV divergence found in \cite{Bern:2008pv} as follows
\be \label{3L4pts}
{M}_4^{\N=(2,2)\text{ L=3}} \;=\; {1\over \epsilon} {5\zeta_3\over (4\pi)^9} \Big ({\kappa\over 2}\Big)^4  \d^6 \left( \sum_{i=1}^4 p_i^{AB} \right) { \delta^{8} \left( \sum_{i=1}^4 q^{A, I}_i \right)  \delta^{8} \left( \sum_{i=1}^4 {\tilde q}^{ {\hat I} }_{i,{\hat A}}\right)   s_{12}\, s_{23}\, s_{34}     } \, ,
\ee
It is proportional to ${M}_4^{\N=(2,2)\text{ tree}}$ and  numerical factors in eq. \rf{3L4pts} are taken from \cite{Bern:2008pv}  where this 3-loop UV divergence was computed.

The presence of a UV divergence in amplitude computations in \cite{Bern:2008pv} and the absence of 1-loop anomalies of a global $USp(4)\times USp(4)$ symmetry discovered in  \cite{Marcus:1985yy} can be considered as a problem. In the earlier case of a 4D $\cN=4$ supergravity anomalies in  \cite{Marcus:1985yy}  and 1-loop U(1) superamplitude anomalies in \cite{Carrasco:2013ypa} were considered as precursors of UV divergences. It was indeed a UV divergence at 4-loop discovered later in \cite{Bern:2013uka}.

Meanwhile in 6D there are no anomalies in  \cite{Marcus:1985yy} and there was no claim of 1-loop superamplitude anomalies: all based on 6D supergravity  \cite{Tanii:1984zk,Bergshoeff:2007ef}
with global $USp(4)\times USp(4)$ symmetry in a symmetric gauge.

Now we have learned that the local $USp(4)\times USp(4)$ symmetry can be gauge-fixed in different gauges. Specifically, 6D supergravity II does not have a global $USp(4)\times USp(4)$ symmetry, same as an Iwasawa gauge of the standard 6D supergravity \cite{Tanii:1984zk,Bergshoeff:2007ef}.  Therefore it may be not totally  surprising  that there is a UV divergence. The theory might have  local $USp(4)\times USp(4)$ anomaly. This would be in agreement with $E_{5(5)}$ anomaly associated with 3-loop UV divergence as discussed in  \cite{Kallosh:2023css,Kallosh:2023dpr,Kallosh:2024lsl}.

\section{Triangular 11D Iwasawa gauge}\label{App:11}

In terms of Fig. \ref{dynkin} this is the parabolic group $P_{\alpha_2}$ when the $\alpha_2$ node is removed.

 In 6D supergravity one can  use a triangle gauge following up the studies of dualities in \cite{Cremmer:1997ct}.
 Here we will show how to impose a triangle gauge  in the model in \cite{Tanii:1984zk} to gauge-fix its local ${SO(5) \times SO(5)}$. This gauge is a part of the family of ${SO(n,n)\over SO(n)\times SO(n)}$ coset spaces where $n=11-D$, i. e. the gauge related to 11D supergravity compactified on a torus $T^{11-D}$ with D$\geq 6$.

It was  found in \cite{Cremmer:1997ct} that the bosonic part of D6 supergravity  in a triangle gauge can be  associated with the dimensional reduction from 11D supergravity on a $T^5$. The relevant scalars are the one defining the torus.

In  \cite{Cremmer:1997ct} the matrix $
\eta =\left(\begin{array}{cc}0 & I \\I & 0\end{array}\right)
$ is left invariant under infinitesimal $SO(5,5)$ transformations
\be
L=\left(\begin{array}{cc}u & v \\\tilde v& -u^T\end{array}\right)
\ee
where $u$ is an arbitrary real matrix, $v=-v^T$, $\tilde v=-\tilde v^T$.

 We take the $U$ matrix in \cite{Tanii:1984zk} and switch to the basis with
$
\eta =\left(\begin{array}{cc}0 & I \\I & 0\end{array}\right)
$
instead of the diagonal in eq. \rf{5.5} so that $\cV \in SO(5,5)$  is related to $U$
\be
U^T= C^T \cV  \, C\, ,  \qquad C^T= {1\over \sqrt 2} \left(\begin{array}{cc}1 & -1 \\1 & 1\end{array}\right)
\ee
${\cV}$
satisfies an $SO(5,5)$  condition as given in  \cite{Cremmer:1997ct} in Appendix C
\be
\cV \, \eta \, \cV ^T=\eta
\label{5.5C}\ee
 The relation between $\cV$ and $U$ is now established.
The triangular gauge, fixing local $SO(5)\times SO(5)$  is
\be
 {\cal V}_{g.f.}=\left(\begin{array}{cc}S & R \\ 0& (S^{-1})^T\end{array}\right)\, , \qquad RS^T + SR^T=0
\ee
Thus, the gauge-fixing condition of $U$ in \cite{Tanii:1984zk} can be given in the form
\be
(U_{gf}^{T})^{triangular}= C^T \left(\begin{array}{cc}S & R \\ 0& (S^{-1})^T\end{array}\right) \, C
\ee
In this gauge there is a  10x10 matrix
\be
{\cal M} = {\cal V}^T {\cal V}= \left(\begin{array}{cc}G_{AB}  & - G_{AC} X^{CB} \\X^{AC} G_{CB} & -X^{AC} G_{CD} X^{DB}\end{array}\right)
\ee
with
\be
(S^T S)_{AB}= G_{AB} \, , \qquad (S^{-1} R)^{AB}= - X^{AB} \, , \qquad X=-X^T
\ee
The ${O(5,5)\over O(5)\times O(5)}$ coset Lagrangian is
\be
{1\over 8} e \Tr  (\partial_\mu {\cal M}^{-1} \partial^\mu {\cal M})= -{1\over 4} G_{AC} G_{BD} (\partial_\mu G^{AB}\partial^\mu G^{CD} + \partial_\mu X^{AB}\partial^\mu G^{CD })
\ee
Thus we have shown here how exactly to fix the triangle $SO(5) \times SO(5)$ gauge in \cite{Tanii:1984zk} to get the 25 moduli in the form
\be
G_{(AB)}, \quad X_{[AB]}\, ,\quad A,B=1,2,3,4,5
\label{torus}\ee
which are the torus $T^5$ moduli.

Note that here the 15 scalars in $G_{AB}$ are on equal footing representing  $T^5$ geometry. But below, in the Iwasawa gauge $P_{\alpha_5}$ related to D+1 dimensions,  we will see that one of 15 is a size of one extra dimension and 14 are coset representatives of ${SL(5, \mathbb{R})\over SO(5,\mathbb{R})}$ coset space inherited from 7D.

In \cite{Cremmer:1997ct} the triangular gauge for 6D maximal supergravity was only given in terms of the 10x10 vielbein, the underlying 16x16 vielbein still has  to fit the one in \cite{Cremmer:1997ct}.
In fact, in \cite{DePol:2000re} there is a proposal how to gauge-fix $V$ in 16x16 representation.

\bibliographystyle{JHEP}
\bibliography{refs}

\end{document}